\documentclass[12pt,prd,nofootinbib]{revtex4-2}

\usepackage{amsmath,amssymb,bm,xcolor}
\usepackage[linktocpage=true]{hyperref}
\newcommand{\U}{{\rm U}}

\newcommand{\diag}{\,{\rm diag}\,}
\newcommand{\link}{\,{\rm Link}\,}

\newcommand{\tf}{\tfrac}

\newcommand{\vev}[1]{\left\langle #1 \right\rangle}
\newcommand{\vevs}[1]{\langle #1 \rangle}

\newcommand{\er}[1]{eq.~\eqref{#1}}
\newcommand{\ers}[1]{eqs.~\eqref{#1}}
\newcommand{\qtq}[1]{\quad\text{#1}\quad}
\newcommand{\ket}[1]{|{#1}\rangle}

\newcommand{\bb}{\mathbb}

\newcommand{\sr}{\sqrt}
\newcommand{\bs}{\boldsymbol}

\newcommand{\fr}{\frac}

\newcommand{\der}{\partial}

\renewcommand{\(}{\left(}
\renewcommand{\)}{\right)}

\newcommand{\dg}{\dagger}

\newcommand{\bmx}{\left(\begin{matrix}}
\newcommand{\emx}{\end{matrix}\right)}
\newcommand{\mtx}[1]{\bmx #1 \emx}

\preprint{KEK-TH-2257}

\allowdisplaybreaks[4]
\begin{document}
\title{Topological mass generation in gapless systems}

\author{Naoki Yamamoto}
\affiliation{Department of Physics, Keio University, Yokohama 223-8522, Japan}
\email{nyama@rk.phys.keio.ac.jp}

\author{Ryo Yokokura}
\affiliation{KEK Theory Center, Tsukuba 305-0801, Japan}
\affiliation{Research and Education Center for Natural Sciences, Keio University, Yokohama 223-8521, Japan}

\email{ryokokur@post.kek.jp}

\begin{abstract}
Mass generation of gauge fields can be universally described
by topological couplings in gapped systems, such as the Abelian Higgs
model in $(3+1)$ dimensions and the Maxwell-Chern-Simons theory in
$(2+1)$ dimensions.
These systems also exhibit the spontaneous breaking of 
higher-form $\bb{Z}_k$
symmetries and topological orders for level $k \geq 2$.

In this paper, we consider topological mass generation in gapless
systems.  
As a paradigmatic example, we study the axion electrodynamics
with level $k$ in $(3+1)$ dimensions in background fields that hosts
both gapped and gapless modes.
We argue that the gapped mode is related to those in fully gapped
systems in lower dimensions via dimensional reduction.  
We show that this system exhibits the spontaneous breaking of a
higher-form $\bb{Z}_k$ symmetry despite the absence of the conventional
topological order.
In the case of the background magnetic field, we also derive the
low-energy effective theory of the gapless mode with the quadratic
dispersion relation and show that it satisfies the chiral anomaly
matching.
\end{abstract}

\maketitle
\tableofcontents
\section{Introduction}
Understanding the origin of the mass is an important question in modern
physics. Among others, the Higgs mechanism provides a mechanism to
explain the mass generation of gauge fields, such as the massive gauge
bosons $W^{\pm}$ and $Z^0$ mediating the weak interaction and massive
photons in superconductivity.  One prototype model of this mechanism is
the Abelian Higgs model, where a $\U(1)$ gauge field becomes massive by
eating a would-be Nambu-Goldstone (NG) boson. The mass generation can
also be described in terms of a low-energy effective theory with the
photon and NG boson, called the Stueckelberg theory~\cite{Stueckelberg:1957zz}.

However, the Higgs mechanism is not a unique mechanism of the mass
generation of gauge fields. In particular, it has been recently shown
that photons can acquire a mass gap in $(3+1)$ dimensions even without
the conventional Higgs mechanism: in the axion electrodynamics in the
presence of background fields, such as a spatially varying axion
field~\cite{Yamamoto:2015maz, Ozaki:2016vwu} or an external magnetic
field~\cite{Sogabe:2019gif,Brauner:2017uiu}, one of the helicity states
of the photons acquires a mass gap, while the other is gapless with the
quadratic dispersion relation. One can ask for a possible universal
description explaining both the Higgs mechanism and this
helicity-dependent mass generation without the Higgs field.

One such possibility is the mass generation due to a topological
coupling of a one-form gauge field and a $(D-2)$-form gauge field in
$D$-dimensional spacetime~\cite{Maldacena:2001ss}.  
Here, ``topological'' means that it does not depend on the metric 
of the spacetime. Examples in this class of theories include the $BF$
theory~\cite{Horowitz:1989ng,Blau:1989dh,Allen:1990gb} with kinetic
terms in $(3+1)$ dimensions~\cite{Cremmer:1973mg,Davis:1988rw},
Maxwell-Chern-Simons theory in $(2+1)$
dimensions~\cite{Binegar:1981gv,Deser:1982vy,Deser:1981wh}, and 
axion electrodynamics in $(1+1)$ 
dimensions~\cite{Witten:1978bc,Aurilia:1980jz}.  
In particular, the
Stueckelberg theory in $D$ dimensions can be dualized to the one-form
and $(D-2)$-form gauge theories with the topological
coupling~\cite{Cremmer:1973mg,Maldacena:2001ss}.

In this paper, we study the axion electrodynamics with level $k$ in
$(3+1)$ dimensions%
\footnote{We will refer to the integer appearing in the coefficient of 
the axion term as the level $k$, similar to the case of the $BF$ theory 
and Maxwell-Chern-Simons theory.} with the background fields above. 
We argue that the helicity-dependent mass generation of photons in this 
theory is related to the topological mass generation in gapped systems 
in lower dimensions via dimensional reduction.  
What is distinct from the conventional topological mass generation is 
that this system also hosts gapless modes; hence, this provides 
an example of ``topological mass generation in gapless systems.'' 
Since this system is gapless, there is no conventional topological order 
that can be seen in gapped systems. 
Nonetheless, we can show that it exhibits the spontaneous breaking of 
a higher-form $\bb{Z}_k$ symmetry similar to the gapped systems 
with topological order.%
\footnote{Generally, the $p$-form symmetries are symmetries under 
transformations of $p$-dimensional extended objects~\cite{Gaiotto:2014kfa}
(see also Refs.~\cite{Batista:2004sc,Pantev:2005zs,Pantev:2005wj,Pantev:2005rh,Nussinov:2006iva,Nussinov:2008aa,Nussinov:2009zz,Banks:2010zn,Distler:2010zg,Nussinov:2011mz,Kapustin:2014gua}).
The symmetry generators are $(D-p-1)$-dimensional topological objects. 
The symmetry transformations are generated by the linking of the symmetry 
generators and the charged objects.}
Moreover, this system also satisfies the chiral 
anomaly matching by the gapless modes. In particular, in the case of 
the external magnetic field, the anomaly matching is satisfied by the
gapless mode with the quadratic dispersion relation (in the transverse
direction) that may be understood as the so-called type-B NG mode
associated with the spontaneous breaking of a one-form symmetry
\cite{Sogabe:2019gif} (see also recent discussion~\cite{Hidaka:2020ucc}).

This paper is organized as follows. 
In Sec.~\ref{sec:gapped}, we review the mass generation mechanisms 
of gauge fields via topological couplings in gapped systems, together with 
concrete examples, such as the $\U (1)$ Abelian Higgs model in $(3+1)$ 
dimensions, $\U (1)$ Maxwell-Chern-Simons theory in $(2+1)$ dimensions, 
and axion electrodynamics in $(1+1)$ dimensions. We highlight the 
spontaneous breaking of higher-form symmetries and its relation to the 
topological order in each of the examples above.  
In Sec.~\ref{sec:gapless}, we study the axion electrodynamics with level 
$k$ in $(3+1)$ dimensions in background fields. We clarify the relation of 
the mass generation mechanism in this gapless system to those in gapped 
systems, the spontaneous breaking of a higher-form $\bb{Z}_k$ symmetry, 
and how the chiral anomaly matching is satisfied at low energy.
Section~\ref{sec:discussion} is devoted to discussions.

Throughout the paper, $a_{\mu}$ denotes a dynamical one-form gauge field, 
$b_{\mu_1...m_{D-2}}$ a dynamical $(D-2)$-form gauge field,
${\bm e}$ a dynamical electric field, ${\bm B}$ an external magnetic field, 
and ${\cal A}_{\mu}, {\cal B}_{\mu \nu}$ background gauge fields. 
We use the ``mostly plus'' metric signature $\eta_{\mu \nu} = {\rm diag} (-1, 1, ..., 1)$
and define the totally antisymmetric tensor $\epsilon^{01...D-1}$ 
so that $\epsilon^{01...D-1} = +1$. 
The $D$-dimensional element on a $D$-dimensional closed subspace, 
${\rm d}S^{\mu_1 \cdots \mu_D}$, is defined such that it is antisymmetric 
with respect to the indices, ${\rm d}S^{\mu_1 \cdots \mu_D} 
= \epsilon^{\mu_1 \cdots \mu_D} {\rm d} S_D$.
We also take the level $k$ to be positive for all the topological couplings 
without loss of generality.

\section{Topological mass generation in gapped systems}
\label{sec:gapped}
In this section, we review the mass generation mechanisms of gauge fields in gapped systems. 
Although the discussions and results of this section are already known in literature, we include 
them to make this paper self-contained, as some of them will be used later in Sec.~\ref{sec:gapless}
as a background. The reader may skip to Sec.~\ref{sec:gapless} and refer back when necessary.

As summarized in Table~\ref{tab:mass}, the mass generation mechanism in gapped systems can be 
described by a topological coupling in arbitrary dimensions~\cite{Maldacena:2001ss}.

\begin{table}[t]
\begin{center}
\begin{tabular}[t]{c|c|c}
\hline\hline
Dimensions & No. of d.o.f. of $a_\mu$
& Topological coupling \\
\hline
$(D-1)+1$ & $D-1$ 
& $\epsilon^{\mu_1...\mu_{D-2}\nu_1\nu_2} b_{\mu_1...\mu_{D-2}} \der_{\nu_1} a_{\nu_2}$
\\
\hline
$3+1$ & 3 
& $\epsilon^{\mu\nu\rho\sigma} b_{\mu\nu} \der_{\rho} a_{\sigma}$
\\
\hline
$2+1$ & 2 
& $\epsilon^{\mu\nu\rho} b_{\mu} \der_{\nu} a_{\rho} $
\\
\hline
$2+1$ & 1 
& $\epsilon^{\mu\nu\rho} a_{\mu} \der_{\nu} a_{\rho} $
\\
\hline
$1+1$ & 1 
& $\epsilon^{\mu\nu} \phi \der_{\mu} a_{\nu} $
\\
\hline
\hline
\end{tabular} 
\end{center}
 \caption{
Mass generation of gauge fields via topological couplings
in various dimensions.
The fields $a_{\mu}$, 
$b_{\mu_1...\mu_{D-2}}$, $b_{\mu\nu}$, $b_\mu $ and $\phi$ denote
a one-form gauge field, a $(D-2)$-form gauge field, a two-form gauge field,
a one-form gauge field, and a scalar field, respectively.
The second column shows the number of massive degrees of freedom 
(d.o.f.) of $a_\mu$ in the presence of the topological coupling.
Note that the number of d.o.f. of massless photons in $D$ dimensions 
is $D-2$.}
\label{tab:mass}
\end{table}

\subsection{$\U(1)$ gauge theories with topological couplings}
\label{sec:topological}
We first consider the mass generation mechanism by the topological coupling 
in $D$-dimensional spacetime~\cite{Maldacena:2001ss}.
We introduce a photon described by a $\U(1)$ one-form gauge field
$a_{\mu}$, whose gauge transformation law is 
$a_{\mu} \to a_{\mu} +\der_{\mu} \lambda$ with a 
$2\pi $ periodic zero-form gauge parameter
$\lambda$.

Here, we assume that the mass dimension of $a_{\mu}$ is one.
Generically, a $p$-form is a $p$th rank antisymmetric tensor.
We normalize $a_{\mu}$ by the flux quantization condition $\int_{{\cal
S}} \frac{1}{2} f_{\mu\nu} {\rm d}S^{\mu \nu} \in 2\pi \bb{Z}$, where
$f_{\mu\nu} = \der_{\mu} a_{\nu} - \der_{\nu} a_{\mu}$ is the field
strength of the gauge field, ${\cal S}$ is a two-dimensional closed surface 
without boundaries, and ${\rm d}S^{\mu \nu}$ is the area element
on ${\cal S}$.

In $D$ dimensions, one can couple the one-form gauge field with 
a $\U(1)$ $(D-2)$-form gauge field $b_{\mu_1...\mu_{D-2}}$
by using the totally antisymmetric tensor $\epsilon^{\mu_1...\mu_D}$. 
We assume that the mass dimension of $b_{\mu_1...\mu_{D-2}}$ 
is $D-2$.
The gauge transformation of $b_{\mu_1...\mu_{D-2}}$ is given by
\begin{equation}
 b_{\mu_1 ... \mu_{D-2}} 
\to 
 b_{\mu_1 ... \mu_{D-2}} 
+ \sum_{\sigma \in S_{D-2}}{\rm sgn}(\sigma) \der_{\mu_{\sigma(1)}} \lambda_{\mu_{\sigma(2)} ... \mu_{\sigma(D-2)}}\,,
\end{equation}
where $\lambda_{\mu_2 ... \mu_{D-2}}$ is a $(D-3)$-form parameter, 
$\sigma$ denotes the permutation of the symmetric group $S_{D-2}$,
and ${\rm sgn}(\sigma) = \pm 1$ for even and odd $\sigma$, respectively.
The normalization of $b_{\mu_1...\mu_{D-2}}$ is 
$\int_{\Sigma_{D-1}} \frac{1}{(D-1)!} h_{\mu_1...\mu_{D-1}} {\rm d}S^{\mu_1 \cdots \mu_{D-1}} \in 2\pi \bb{Z}$, 
where $h_{\mu_1...\mu_{D-1}}$ is the field strength
\begin{equation}
 h_{\mu_1...\mu_{D-1}} 
= \ \sum_{\sigma \in S_{D-1}}{\rm sgn}(\sigma) \der_{\mu_{\sigma(1)}} b_{\mu_{\sigma(2)} ... \mu_{\sigma(D-1)}}\,,
\end{equation}
$\Sigma_{D-1}$ is a $(D-1)$-dimensional closed subspace without boundaries,
and ${\rm d}S^{\mu_1 \cdots \mu_{D-1}}$ is the $(D-1)$-dimensional element on $\Sigma_{D-1}$.

We start with the following action:
\begin{equation}
\begin{split}
 S_{\rm top}
& = -\int {\rm d}^D x
\(
\fr{1}{8\pi^2 v_D^2 (D-1)!} | h_{\mu_1...\mu_{D-1}} |^2
 +\fr{1}{4e_D^2} |f_{\mu\nu}|^2\)
\\
&
\quad
+\fr{k}{2\pi}
\int {\rm d}^D x
\fr{\epsilon^{\mu_1...\mu_{D-2} \nu_1\nu_2}}{(D-2)!\cdot 2!} 
b_{\mu_1...\mu_{D-2}}f_{\nu_1\nu_2}\,,
\end{split}
\label{topological_D}
\end{equation}
where $v_D$ is some constant with mass dimension $\fr{D -2}{2}$ and
$e_D$ is a coupling constant with mass dimension $ \fr{4- D}{2}$.
The first line describes the kinetic terms of $a_{\mu}$ and $b_{\mu_1...\mu_{D-2}}$, 
and the second line is the topological coupling.
Note that this coupling does not depend on the metric of spacetime
since the vector indices are contracted with $\epsilon^{\mu_1...\mu_D}$, 
and hence, it is topological.%
\footnote{More explicitly, this can be seen as follows. 
As a generic vector field $V_{\mu}(x)$ transforms under a 
coordinate transformation $x \rightarrow x'(x)$ as
$V_{\mu}(x) = \fr{\der x'^{\rho}}{\der x^{\mu}} V'_{\rho}(x')$, 
we have
\begin{align}
\epsilon^{\mu_1...\mu_{D-2} \nu_1\nu_2} b_{\mu_1...\mu_{D-2}}f_{\nu_1\nu_2}
& 
= \epsilon^{\mu_1...\mu_{D-2} \nu_1\nu_2} 
\fr{\der x'^{\rho_1}}{\der x^{\mu_1}} ...  \fr{\der x'^{\rho_{D-2}}}{\der x^{\mu_{D-2}}}
\fr{\der x'^{\lambda_1}}{\der x^{\nu_1}} \fr{\der x'^{\lambda_2}}{\der x^{\nu_2}}
b'_{\rho_1...\rho_{D-2}}f'_{\lambda_1\lambda_2}
\nonumber \\
& 
= \det \left(\fr{\der x'}{\der x} \right)
\epsilon^{\rho_1...\rho_{D-2} \lambda_1\lambda_2} 
b'_{\rho_1...\rho_{D-2}}f'_{\lambda_1\lambda_2}\,.
\nonumber 
\end{align}
Combined with the relation ${\rm d}^D x' = \det \left(\fr{\der x'}{\der x} \right) {\rm d}^D x$,
we arrive at
\begin{equation}
{\rm d}^D x \ \epsilon^{\mu_1...\mu_{D-2} \nu_1\nu_2} b_{\mu_1...\mu_{D-2}}f_{\nu_1\nu_2}
= {\rm d}^D x' \ \epsilon^{\rho_1...\rho_{D-2} \lambda_1\lambda_2} 
b'_{\rho_1...\rho_{D-2}}f'_{\lambda_1\lambda_2}\,. \nonumber
\end{equation}
Hence, the last term in Eq.~(\ref{topological_D}) is invariant under
general coordinate transformations without the metric.}
The topological term is invariant under the gauge transformations 
of $a_{\mu}$ and $b_{\mu_1...\mu_{D-2}}$.
The constant $k$ is restricted as $k \in \bb{Z}$ by the invariance 
under large gauge transformations of $a_{\mu}$ and $b_{\mu_1...\mu_{D-2}}$.

Since the topological term is quadratic with the first-order derivative, 
it generates the masses of $a_{\mu}$ and $b_{\mu_1...\mu_{D-2}}$.
The equations of motion of $a_{\mu}$ and $b_{\mu_1...\mu_{D-2}}$ are
\begin{gather}
\label{a-EOM_D}
  \fr{1}{e_D^2} \der_{\nu} f^{\nu \mu} - 
\fr{k}{2\pi} \fr{\epsilon^{\nu_1...\nu_{D-2} \rho\mu}}{(D-2)!}
 \der_{\rho} b_{\nu_1...\nu_{D-2}} = 0\,, \\
 \label{b-EOM_D}
  \fr{1}{4\pi^2 v_D^2} \der_{\rho} h^{\rho\mu_1...\mu_{D-2}} +
\fr{k}{2\pi} \epsilon^{\mu_1...\mu_{D-2} \nu \rho} \der_{\nu} a_{\rho} = 0 \,,
\end{gather}
respectively.
By substituting the former into the latter and vice versa, 
we have
\begin{equation}
(\der^2 - \Delta_{D}^2) f_{\mu\nu} =0, \qquad
(\der^2 - \Delta_{D}^2) h_{\mu_1...\mu_{D-2}} =0,
\end{equation}
where $\der^2 \equiv \der_\rho \der^\rho$ and 
\begin{equation}
\Delta_{D}^2 = (k e_D v_D)^2,
\end{equation}
which shows that $f_{\mu\nu}$ and $h_{\mu_1...\mu_{D-2}}$ 
are both massive.

In order to count the physical degrees of freedom, we locally solve 
the equations of motion in \ers{a-EOM_D} and \eqref{b-EOM_D} as
\begin{gather}
\label{a-sol_D}
  \fr{1}{e_D^2} f^{\rho \mu} - 
\fr{k}{2\pi}  \fr{\epsilon^{\nu_1...\nu_{D-2} \rho \mu}}{(D-2)!} 
[b_{\nu_1...\nu_{D-2}} - (D-2) \der_{\nu_1} \bar b_{\nu_2...\nu_{D-2}}] = 0\,, \\
\label{b-sol_D}
  \fr{1}{4\pi^2 v_D^2} h^{\rho\mu_1...\mu_{D-2}} +
\fr{k}{2\pi} \epsilon^{\mu_1...\mu_{D-2}\rho\sigma} (a_{\sigma} -\der_{\sigma} \bar a) = 0\,.
\end{gather}
Here, 
the $(D-3)$- and zero-form fields $\bar b_{\nu_2...\nu_{D-2}}$ 
and $\bar a$ are possible ambiguities of the solutions,
and they can be absorbed into 
$b_{\mu_1...\mu_{D-2}}$ and $a_{\mu}$ by gauge fixing.
Combining the solutions in \ers{a-sol_D} and \eqref{b-sol_D} 
with the Bianchi identities leads to the constraints on the gauge fields, 
\begin{equation}
\fr{k}{2\pi} 
\der^{\nu_1} b_{\nu_1...\nu_{D-2}} = 0\,, 
\qquad
\fr{k}{2\pi} \der^{\sigma} a_{\sigma} = 0 \,,
\label{constraint_D}
\end{equation}
which provide $\fr{(D-1)(D-2)}{2}$ and $1$ constraints, respectively.
Therefore, both $a_\mu$ and $b_{\mu_1...\mu_{D-2}}$ have 
$D-1 = \fr{D(D-1)}{2} - \fr{(D-1)(D-2)}{2}$ degrees of freedom.
Furthermore, $b_{\mu_1...\mu_{D-2}}$ and $a_\mu$ are related to 
each other by the equations of motion, and the physical degrees of 
freedom of this system is $D-1$.

\subsection{Dual Stueckelberg action}
\label{sec:dual}
It is known that the action with the topological coupling in \er{topological_D}
is dual to the so-called Stueckelberg action~\cite{Stueckelberg:1957zz},
\begin{equation}
 S_{\rm St}
 = -\int {\rm d}^D x
\(
\fr{v_D^2}{2} |\der_{\mu} \chi  - ka_{\mu}|^2 
+\fr{1}{4e_D^2} |f_{\mu\nu}|^2
\)\,,
\label{Stuckelberg}
\end{equation}
where $\chi$ is a $2\pi$ periodic scalar field. 
The gauge transformation of $\chi$ is $\chi \to \chi + k\lambda $ with 
$a_{\mu} \to a_{\mu} + \der_{\mu} \lambda$. One can impose the gauge-fixing 
condition $\chi =0$ (unitary gauge) by using the gauge freedom of 
$a_{\mu}$. In this gauge, the term $|\der_{\mu} \chi  - ka_{\mu}|^2$ 
becomes the mass term of $a_{\mu}$.

The Stueckelberg action describes a massive photon with $(D-1)$ polarizations.
The equations of motion of $a_{\mu}$ and $\chi$ are given by
\begin{gather}
  v_D^2 k (\der_{\mu} \chi - ka_{\mu}) + \fr{1}{e_D^2} \der^{\nu} f_{\nu \mu} = 0\,, \\
  v_D^2 \der^{\mu} (\der_{\mu} \chi- ka_{\mu}) =0\,,
\end{gather}
respectively.
Under the gauge condition $\chi =0$, 
the equation of motion of $a_{\mu}$ reduces to
\begin{equation}
(\der^2 - \Delta_{D}^2) a_{\mu}  =0,
\end{equation}
with one constraint
\begin{equation}
  k v_D^2 \der^{\mu} a_{\mu} =0.
\end{equation}
The latter corresponds to the constraint in \er{constraint_D}.

These equations show the presence of massive excitations with 
three polarizations. More explicitly, for the plane wave ansatz
$a_{\mu} = \epsilon_{\mu} (p) {\rm e}^{- {\rm i}\omega t + {\rm i}\bs{p}\cdot \bs{x}}$ 
with $p^{\mu} = (\omega, \bs{p})$ the $D$ momentum, we have
\begin{gather}
 (\omega^2 - \bs{p}^2 - \Delta_{D}^2)  \epsilon_{\mu} (p) =0, \\
  p^{\mu} \epsilon_{\mu} (p) =0.
\end{gather}
Therefore, it has the mass gap, and one of the polarizations, $\epsilon_0(p)$, 
vanishes in the rest frame of the massive gauge field.

We now see that the Stueckelberg action is dual to the action 
with the topological coupling~\cite{Cremmer:1973mg,Maldacena:2001ss}.
In other words, they are (classically) equivalent to each other.
To see this, we introduce the following action:
\begin{equation}
\begin{split}
  S'_{\rm St}
& =-  \int {\rm d}^D x
\( \fr{v_D^2}{2} |w_{\mu}|^2
 + \fr{1}{4e_D^2} |f_{\mu\nu}|^2\) 
\\
&
\quad
+\int {\rm d}^D x
\fr{(-1)^{D-2}}{2\pi} \fr{\epsilon^{\mu_1... \mu_{D-1} \nu}}{(D-1)!}
 h_{\mu_1...\mu_{D-1}} (w_{\nu} - \der_{\nu} \chi - k a_{\nu}) \,,
\end{split}
\label{S'}
\end{equation}
where $w_{\mu}$ and $h_{\mu_1...\mu_{D-1}}$ are one- and 
$(D-1)$-form fields, respectively.
Using the equation of motion of $h_{\mu_1...\mu_{D-1}}$, we can 
recover the original action \eqref{Stuckelberg}.
Instead, we can dualize the action by eliminating $\chi$.
First, the equation of motion of $\chi$ is given by
$\epsilon^{\mu_1 ...\mu_{D-1}\nu} \der_{\nu}  h_{\mu_1...\mu_{D-1}} =0$.
Then, the $(D-1)$-form field $ h_{\mu_1...\mu_{D-1}}$ can be locally 
expressed by a $(D-2)$-form gauge field $b_{\mu_1...\mu_{D-2}}$ as
\begin{equation}
 h_{\mu_1...\mu_{D-1}} 
= \ \sum_{\sigma \in S_{D-1}}{\rm sgn}(\sigma) \der_{\mu_{\sigma(1)}} b_{\mu_{\sigma(2)} ... \mu_{\sigma(D-1)}}\,.
\end{equation}
The solution is invariant under the gauge transformation of $b_{\mu_1...\mu_{D-2}}$,
\begin{equation}
 b_{\mu_1 ... \mu_{D-2}} 
\to 
 b_{\mu_1 ... \mu_{D-2}} 
+
\sum_{\sigma \in S_{D-2}}{\rm sgn}(\sigma) \der_{\mu_{\sigma(1)}} \lambda_{\mu_{\sigma(2)} ... \mu_{\sigma(D-2)}}\,,
\end{equation}
where $\lambda_{\mu_2 ... \mu_{D-2}}$ is a $(D-3)$-form parameter.
Second, the equation of motion of $w_{\mu}$ is
\begin{equation}
 v_D^2 w^{\nu} = 
\fr{(-1)^{D-2}}{2\pi}\fr{\epsilon^{\mu_1...\mu_{D-1}\nu}}{(D-1)!} h_{\mu_1...\mu_{D-1}}\,.
\end{equation}
Substituting these equations into the action \eqref{S'}, 
we get the action with the topological term in \er{topological_D}.

\subsection{Topological order and 
spontaneous breaking of higher-form $\bb{Z}_k$ symmetries}
\label{to}
Here, we comment on the fact that the system above with $D\geq 3$ 
has the so-called topological order~\cite{Wen:1989zg,Wen:1989iv,Hansson:2004wca} 
and that it exhibits the so-called type-B spontaneous symmetry breaking 
(SSB)~\cite{Nielsen:1975hm,Watanabe:2011ec,Watanabe:2012hr,Hidaka:2012ym} 
of $\bb{Z}_k$ one- and $(D-2)$-form global symmetries~\cite{Gaiotto:2014kfa}.%
\footnote{Generally, type-A and type-B SSB are characterized by the conditions 
that commutation relations of the broken symmetry generators are zero 
and nonzero, respectively, whichever for continuous and discrete symmetries.}

In order to see the topological nature of the system, we consider the low-energy 
limit, where the kinetic terms of the one- and $(D-2)$-form gauge fields
are negligible. Consequently, the theory can be described by the topological action
\begin{equation}
 S_{BF}
 = \fr{k}{2\pi}
 \int {\rm d}^D x
 \fr{\epsilon^{\mu_1...\mu_{D-2}\nu_1\nu_2}}{(D-2)!\cdot 2!} 
b_{\mu_1...\mu_{D-2}}f_{\nu_1\nu_2}\,.
\label{BF} 
\end{equation}
This is the so-called $BF$ theory~\cite{Horowitz:1989ng,Horowitz:1989km}.
In this effective theory, there is no local observable since the field strengths 
of the gauge fields are zero by the equations of motion, 
$f_{\rho \sigma} =0$ and $h_{\mu_1...\mu_{D-1}} =0$.
However, there are nonlocal observables, i.e., 
the Wilson loop and the vortex world volume,
\begin{equation}
 W({\cal C}) = {\rm e}^{{\rm i}\int_{\cal C} a_{\mu} {\rm d}x^{\mu} },
\qtq{and}
 V(\Sigma_{D-2}) 
= {\rm e}^{{\rm i} \int_{\Sigma_{D-2}} \frac{1}{(D-2)!}
b_{\mu_1...\mu_{D-2}} {\rm d}S^{\mu_1 \cdots \mu_{D-2}} },
\end{equation}
respectively.
Here, ${\cal C}$ is a one-dimensional closed loop, $\Sigma_{D-2}$ is 
a $(D-2)$-dimensional closed subspace without boundaries, and 
${\rm d}S^{\mu_1 \cdots \mu_{D-2}}$ is the $(D-2)$-dimensional 
element on $\Sigma_{D-2}$.
Physically, the Wilson loop is a world line of a probe particle with 
a unit charge and $ V(\Sigma_{D-2}) $ is a world volume of 
a codimension 2 quantized magnetic vortex, which is 
a higher-dimensional generalization of Abrikosov-Nielsen-Olesen 
vortex in $(3+1)$ dimensions~\cite{Abrikosov:1956sx,Nielsen:1973cs}.

The topological excitations can be seen by the correlation function
\begin{equation}
 \vevs{W({\cal C})V(\Sigma_{D-2}) } 
 \equiv {\cal N} \int {\cal D} a_{\mu} {\cal D}b_{\nu_1...\nu_{D-2}} 
W({\cal C}) V(\Sigma_{D-2}) {\rm e}^{{\rm i} S_{BF}}
 = {\rm e}^{\fr{2\pi {\rm i}}{k} \link ({\cal C},\Sigma_{D-2})}\,,
\label{correlation}
\end{equation}
where ${\cal N}$ is a normalization factor such that $\vevs{1} =1$,
and ``$\link ({\cal C},\Sigma_{D-2})$'' denotes the linking number of 
${\cal C}$ and $\Sigma_{D-2}$.

Equation~(\ref{correlation}) can be derived by integrating out 
the Wilson loop and the vortex world volume~\cite{Horowitz:1989km,Oda:1989tq} 
(see also recent Refs.~\cite{Chen:2015gma,Hidaka:2019jtv}).
To integrate out the Wilson loop first, we rewrite the line integral 
$\int_{\cal C} a_{\mu} {\rm d}x^{\mu}$ by the Stokes' theorem as
\begin{equation}
\label{Stokes}
\int_{\cal C} a_{\mu} {\rm d}x^{\mu} = 
\int_{\der \cal S_C} a_{\mu} {\rm d}x^{\mu} = 
\int_{\cal S_C} \frac{1}{2} f_{\mu\nu} {\rm d}S^{\mu \nu}\,.
\end{equation}
Here, ${\cal S_C}$ is a two-dimensional surface whose 
boundary is ${\cal C}$, $\der {\cal S_C} = {\cal C}$ with
$\der$ being the boundary operator.
We then rewrite the surface integral 
$\int_{\cal S_C} \frac{1}{2} f_{\mu\nu} {\rm d}S^{\mu \nu}$ into 
the spacetime integral by using the delta function as
\begin{equation}
\begin{split}
 \int_{\cal S_C} \frac{1}{2} f_{\mu\nu} (y) {\rm d}S^{\mu \nu}(y)
 &= 
\int {\rm d}^D x 
\int_{\cal S_C}  {\rm d}S^{\mu \nu}(y)
\frac{1}{2}
 \delta^D (x-y)
 f_{\mu\nu} (x) 
\\
& = \int {\rm d}^D x \frac{1}{2} f_{\mu\nu} (x) J^{\mu\nu}(x;{\cal S_C})\,,
\label{S->V}
\end{split}
\end{equation}
where $ J^{\mu\nu}(x;{\cal S_C})$ is defined by 
\begin{equation}
 J^{\mu\nu}(x;{\cal S_C}) =  \int_{\cal S_C} \delta^D (x-y) {\rm d}S^{\mu \nu}(y)\,.
\label{delta_current}
\end{equation}
In the following we abbreviate $ J^{\mu\nu}(x;{\cal S_C})$ to 
$ J^{\mu\nu}({\cal S_C})$.

Using Eqs.~(\ref{Stokes}) and (\ref{S->V}), the topological action 
and the line integral in \er{correlation} can be written as
\begin{equation}
\begin{split}
S_{BF} + \int_{\cal C} a_{\mu} {\rm d}x^{\mu} 
= 
\fr{k}{2\pi}
\int {\rm d}^D x
\fr{1}{2!}\(\fr{1}{(D-2)!} 
\epsilon^{\mu_1...\mu_{D-2} \nu_1\nu_2} b_{\mu_1...\mu_{D-2}} + 
\fr{2\pi }{k} J^{\nu_1\nu_2}({\cal S_C})\) f_{\nu_1\nu_2}\,.   
\end{split}
\end{equation}
The Wilson loop can now be integrated out by the redefinition
\begin{equation}
 b_{\mu_1...\mu_{D-2}} 
\to 
 b_{\mu_1...\mu_{D-2}}
+
\fr{\pi }{k} \epsilon_{\mu_1...\mu_{D-2} \nu_1\nu_2} 
J^{\nu_1\nu_2}({\cal S_C})\,.
\end{equation}
Meanwhile, this redefinition transforms the vortex world volume 
in \er{correlation} as
\begin{equation}
\begin{split}
&\int_{\Sigma_{D-2}} 
b_{\mu_1...\mu_{D-2}} {\rm d}S^{\mu_1 \cdots \mu_{D-2}} 
\\
&
\to 
\int_{\Sigma_{D-2}} 
b_{\mu_1...\mu_{D-2}} {\rm d}S^{\mu_1 \cdots \mu_{D-2}} 
+
\int_{\Sigma_{D-2}} 
\fr{\pi }{k} \epsilon_{\mu_1...\mu_{D-2} \nu_1\nu_2} 
J^{\nu_1\nu_2}({\cal S_C})
{\rm d}S^{\mu_1 \cdots \mu_{D-2}} \,.
\end{split}
\end{equation}
The integral of the last term leads to the intersection number of 
${\cal S_C}$ and 
$\Sigma_{D-2}$, which is equal to the linking number of 
${\cal C}$ and $\Sigma_{D-2}$:%
\footnote{We can relate the linking number to 
a $D$-dimensional Gauss linking number by the relation of 
the delta function and Green's function~\cite{Oda:1989tq}.}
\begin{equation}
\int_{\Sigma_{D-2}} \fr{1}{(D-2)! \cdot 2!}
\epsilon_{\mu_1...\mu_{D-2} \nu_1\nu_2} J^{\nu_1\nu_2}({\cal S_C})
{\rm d}S^{\mu_1 \cdots \mu_{D-2}} 
=\link({\cal C}, \Sigma_{D-2})\,.
\end{equation}
Therefore, the correlation function is  
\begin{equation}
 \vevs{W({\cal C})V(\Sigma_{D-2}) } 
 = {\rm e}^{\fr{2\pi {\rm i} }{k}\link({\cal C}, \Sigma_{D-2})}
\vevs{V(\Sigma_{D-2})}\,.
\label{correlation_V}
\end{equation}
The vortex world volume can be similarly integrated out.
In this procedure, the redefinition of $a_{\mu}$ does not give an additional 
contribution, since the Wilson loop has been already integrated out.
Therefore, we have 
\begin{equation}
 \vevs{V(\Sigma_{D-2})} = 1,
\label{V}
\end{equation}
and we arrive at \er{correlation}.

Of course, we can first integrate out 
the vortex world volume, then integrate out the 
Wilson loop. In this case, we have
\begin{equation}
  \vevs{W({\cal C})V(\Sigma_{D-2}) } 
 = {\rm e}^{\fr{2\pi {\rm i} }{k}\link({\cal C}, \Sigma_{D-2})}
\vevs{W({\cal C})},
\label{correlation_W}
\end{equation}
and 
\begin{equation}
 \vevs{W({\cal C})} = 1. 
\label{W}
\end{equation}
Therefore, we have the same result as \er{correlation}.

The topological feature of the correlation function (\ref{correlation})
is that it depends only on the linking number.
This is the topologically ordered phase~\cite{Wen:1989zg,Wen:1989iv,Hansson:2004wca}, 
which can generally be characterized by the condition that the correlation function 
of the spatially and temporally extended topological objects have a nonzero 
fractional phase if they are linked to each other.

The fractional phase due to the linking also leads to the ground-state 
degeneracy on a compact spatial manifold. 
To see this, it is convenient to switch to the operator formalism.
Let us consider the system on the spacetime with a $(D-1)$-dimensional 
spatial manifold $M_{D-1}$ with nontrivial topology 
such that both of the Wilson loop and vortex world volume can 
topologically wrap subspaces of $M_{D-1}$.
One of the simplest choice may be $M_{D-1} = S^{D-2} \times S^1$, 
where $S^{D-2}$ and $S^1$ are a $(D-2)$-dimensional sphere and 
a circle, respectively.
Since $V(\Sigma_{D-2})$ is a topological object, the action of 
$V(\Sigma_{D-2})$ on the ground state $\ket{\Omega}$ 
does not change the energy of the system.
Therefore, we can choose the ground state as an eigenstate of
$V(\Sigma_{D-2})$ with the eigenvalue ${\rm e}^{{\rm i}\theta}$.
Meanwhile, the Wilson loop $W({\cal C})$ does not change 
the energy of the system. Thus, we have another ground state 
$\ket{\Omega'} = W({\cal C})\ket{\Omega}$.
We can show that $\ket{\Omega'}$ and $\ket{\Omega}$ are orthogonal to
each other. In fact, the inner product $\vevs{\Omega|\Omega'}$ can be 
evaluated as
\begin{equation}
\begin{split}
 \vevs{\Omega|\Omega'} 
&= 
\vevs{\Omega|W({\cal C})|\Omega}= 
\vevs{\Omega|V^{-1}(\Sigma_{D-2})W({\cal C}) V(\Sigma_{D-2})|\Omega}
= 
{\rm e}^{\fr{2\pi {\rm i}}{k}}\vevs{\Omega|W({\cal C}) |\Omega}
\\
&
= 
{\rm e}^{\fr{2\pi {\rm i}}{k}}\vevs{\Omega|\Omega'}.
\end{split}
\end{equation}
and so $ \vevs{\Omega|\Omega'} = 0$ for $k \geq 2$.
Here, we used \er{correlation_W}, which, in the operator formalism, 
leads to the equal-time commutation relation
\begin{equation}
V^{-1}(\Sigma_{D-2})W({\cal C}) V(\Sigma_{D-2})
= 
{\rm e}^{\fr{2\pi {\rm i}}{k}} W({\cal C}), 
\end{equation}
if $\Sigma_{D-2}$ and ${\cal C}$ has intersection number on 
$M_{D-1}$. Since $W({\cal C})^{n}$ ($n= 0,..., k-1$) leads to 
a different phase, there are $k$ degenerate ground states.

In the viewpoint of global symmetries, the correlation function 
(\ref{correlation}) shows that there are spontaneously broken
$\bb{Z}_k$ one- and $(D-2)$-form global symmetries for $k \geq 2$,
and the symmetry breaking pattern is classified as type B.
To see this, recall that charged objects under the one- and 
$(D-2)$-form symmetries are $W({\cal C})$ and $V(\Sigma_{D-2})$, 
respectively. The relations (\ref{correlation_W}) and (\ref{correlation_V})
mean that the symmetry generators of the one- and $(D-2)$-form 
symmetries are $V(\Sigma_{D-2})$ and $W({\cal C})$, respectively.
Both of these symmetry transformations are parametrized by the discrete 
group $\bb{Z}_k$, since $\vevs{W({\cal C})^k V(\Sigma_{D-2})} = 1$
and $\vevs{W({\cal C}) V(\Sigma_{D-2})^k} = 1$.
Then, the nonzero vacuum expectation values (VEVs) in \ers{W} and \eqref{V} 
show that the $\bb{Z}_k$ one- and $(D-2)$-form symmetries 
are spontaneously broken. 
In addition, the correlation function (\ref{correlation}) shows that 
the symmetry breaking pattern is type B, where broken symmetry 
generators are also charged objects.

It should be remarked that, for the existence of the topologically ordered 
phase, the condition that the system has a mass gap is essential.
If the system is gapless, an infinitesimal deformation of the Wilson loop 
or vortex world volume gives rise to excited states of the one-form or 
$(D-2)$-form gauge field even in the low-energy limit. 
In this case, the states $W({\cal C}) \ket{\Omega}$ 
and $V(\Sigma_{D-2}) \ket{\Omega}$ cannot be ground states. Then, 
there is no ground-state degeneracy even if the system is put on a 
topologically nontrivial spatial manifold, and there is no topological order.
In Sec.~\ref{sec:gapless}, we will consider a gapless system with 
the type-B SSB, which may not be a topologically ordered phase.

\subsection{Examples}
Let us now review concrete examples of the mass generation mechanism above.

\subsubsection{$\U(1)$ Abelian Higgs model in $(3+1)$ dimensions}
The first example is the $\U(1)$ Abelian Higgs model in $(3+1)$ dimensions.
In the dual theory of the Abelian Higgs model, the mass of the gauge field 
is generated by a topological coupling with a 
two-form gauge field~\cite{Davis:1988rw,Lee:1993ty}.

We consider a Higgs field $\Phi$ in a $\U(1)$ gauge theory.
The $\U(1)$ charge of $\Phi$ is $k$, $\Phi \to {\rm e}^{{\rm i}k\lambda} \Phi$
with $a_{\mu} \to a_{\nu} + \der_{\mu}\lambda$, where $a_{\mu}$ and $\lambda$ 
are a $\U(1)$ one-form gauge field and a zero-form gauge parameter,
respectively. We introduce the action
\begin{equation}
  S_{\rm AH}
 = -\int {\rm d}^4 x
\(
\fr{v^2}{2} |\der_{\mu} \Phi  - {\rm i} k a_{\mu} \Phi|^2 
+V(|\Phi|)
+\fr{1}{4e^2} |f_{\mu\nu}|^2
\)\,,
\end{equation}
where $V(|\Phi|)$ is a potential of the Higgs field such that the 
Higgs field develops a nonzero VEV $\vevs{\Phi} = \fr{v}{\sr{2}}$.

Well below the mass of the radial excitation of the Higgs field, 
the low-energy effective action of this theory can be 
written by the Stueckelberg action
\begin{equation}
 S_{\rm AH,eff}
 = -\int {\rm d}^4 x
\(
\fr{v^2}{2} |\der_{\mu} \chi  - ka_{\mu}|^2 
+\fr{1}{4e^2} |f_{\mu\nu}|^2
\).
\label{AH_eff}
\end{equation}
In $(3+1)$ dimensions, the Stueckelberg action describes 
a massive photon with three polarizations.

One can dualize the effective action (\ref{AH_eff}) to 
a two-form gauge theory.
By the same procedure explained in Sec.~\ref{sec:dual}, 
we arrive at the following action with a topological coupling 
of the one- and two-form gauge fields:
\begin{equation}
 S_{\rm AH, dual}
 =  \int {\rm d}^4 x
\(
-\fr{1}{8\pi^2 v^2\cdot 3!} |h_{\mu\nu\rho}|^2
 -  \fr{1}{4e^2} |f_{\mu\nu}|^2
+\fr{k}{8\pi} \epsilon^{\mu\nu\rho\sigma} b_{\mu\nu}f_{\rho\sigma} 
\)\,,
\label{AH_dual}
\end{equation}
where $b_{\mu\nu}$ is the two-form gauge field and 
$h_{\mu\nu\rho} = \der_{\mu} b_{\nu \rho}+\der_{\nu} b_{\rho\mu}+ \der_{\rho} b_{\mu\nu}$
is the field strength. The topological term give rises to the masses 
of $a_{\mu}$ and $b_{\mu\nu}$.

The phase of the Abelian Higgs theory can be classified 
by the type-B SSB of $\bb{Z}_k$ one- and two-form symmetries.
Here, the charged objects under the one- and two-form symmetries 
are the Wilson loop of a probe particle $W({\cal C})$ and the 
world surface of a quantized magnetic vortex string $V({\cal S})$, with
a two-dimensional closed world surface ${\cal S}$. 
This is the Abrikosov-Nielsen-Olesen 
vortex string~\cite{Abrikosov:1956sx,Nielsen:1973cs}.
The generators of the one- and two-form symmetries are
$V({\cal S})$ and $W({\cal C})$, respectively.
Explicitly, the correlation function of $W({\cal C})$ 
and $V({\cal S})$ is evaluated by the $BF$ theory as
\begin{align}
 \vevs{W({\cal C})V({\cal S})}
&= {\rm e}^{\fr{2\pi {\rm i}}{k}\link ({\cal C,S})} \vevs{V({\cal S})}
\nonumber \\
&= {\rm e}^{\fr{2\pi {\rm i}}{k}\link ({\cal C,S})} \vevs{W({\cal C})}
\nonumber \\
& ={\rm e}^{\fr{2\pi {\rm i}}{k}\link ({\cal C,S})} . 
\end{align}
The first line means that $W({\cal C})$ and $V({\cal S})$ 
are the symmetry generator and the charged object
under the $\bb{Z}_k$ two-form symmetry, respectively.
The second line means that $V({\cal S})$ and $W({\cal C})$ 
are the symmetry generator and the charged object
under the $\bb{Z}_k$ one-form symmetry, respectively.
The third line shows that both of the $\bb{Z}_k$ one- and two-form 
symmetries are broken spontaneously, and 
the correlation functions of the broken symmetry generators is 
finite; thus, this can be classified as the type-B SSB.
The Higgs phase is further classified as a topologically ordered 
phase~\cite{Hansson:2004wca}, since both of the extended 
charged objects obey the nontrivial braiding statistics.

\subsubsection{Topological coupling with two one-form gauge fields}

We then consider the $(2+1)$-dimensional version of the 
action (\ref{topological_D}).
The action is given by 
\begin{equation}
\begin{split}
  S_{\rm 3D} 
&= -\int {\rm d}^3x 
\(\fr{1}{4e_{\rm 3D}^2} |f_{\mu\nu}|^2 +
\fr{1}{16\pi^2 v_{\rm 3D}^2} |h_{\mu\nu}|^2 
- \fr{k}{2\pi}  \epsilon^{\mu\nu\rho} b_{\mu} \der_{\nu} a_{\rho} \) \,,
\end{split}
\end{equation}
where $e_{\rm 3D}$ is a coupling constant with mass dimension $1/2$
and $v_{\rm 3D}$ is a constant with mass dimension $-1/2$.
The action gives one massive photon with two polarizations.

This theory is characterized by the type-B SSB of two 
$\bb{Z}_k$ one-form global symmetries.
The charged objects are Wilson loops 
${\rm e}^{{\rm i}\int_{\cal C}a_{\mu} {\rm d}x^{\mu}}$ 
and ${\rm e}^{{\rm i}\int_{\cal C} b_{\mu} {\rm d}x^{\mu}}$.  
Again, the correlation function is
\begin{equation}
 \vevs{
{\rm e}^{{\rm i}\int_{\cal C} a_{\mu} {\rm d}x^{\mu}}
{\rm e}^{{\rm i}\int_{\cal C'} b_{\mu} {\rm d}x^{\mu}}}
= 
{\rm e}^{\fr{2\pi {\rm i}}{k}\link ({\cal C,C'})} \vevs{
{\rm e}^{{\rm i}\int_{\cal C'} b_{\mu} {\rm d}x^{\mu}}}
= 
{\rm e}^{\fr{2\pi {\rm i}}{k}\link ({\cal C,C'})} \vevs{
{\rm e}^{{\rm i}\int_{\cal C} a_{\mu} {\rm d}x^{\mu}}
}
= 
{\rm e}^{\fr{2\pi {\rm i}}{k}\link ({\cal C,C'})}.
\end{equation}
These relations show that the generators of the two $\bb{Z}_k$ one-form
symmetries are ${\rm e}^{{\rm i}\int_{\cal C} a_{\mu} {\rm d}x^{\mu}}$
and ${\rm e}^{{\rm i}\int_{\cal C'} b_{\mu} {\rm d}x^{\mu}}$, and the
charged objects are ${\rm e}^{{\rm i}\int_{\cal C'} b_{\mu} {\rm
d}x^{\mu}}$ and ${\rm e}^{{\rm i}\int_{\cal C} a_{\mu} {\rm d}x^{\mu}}$,
respectively.  
Furthermore, both of them exhibit the type-B SSB, and so it is a
topologically ordered phase.
Note that this action is dual to the $(2+1)$-dimensional Stueckelberg
action
\begin{equation}
S_{{\rm 3D, St}}
 = -\int {\rm d}^3x 
\(
\fr{v_{\rm 3D}^2}{2} |\der_{\mu} \chi  - ka_{\mu}|^2 
+\fr{1}{4e_{\rm 3D}^2} |f_{\mu\nu}|^2
\)\,,
\end{equation}
which is the low-energy effective theory of the $(2+1)$-dimensional Abelian
Higgs model.

\subsubsection{$\U(1)$ Maxwell-Chern-Simons theory in $(2+1)$ dimensions}
\label{sec:MCS}
In the above discussion, we have introduced two independent one-form gauge
fields. However, one can introduce a topological term by using a single 
one-form gauge field in $(2+1)$ dimensions, called the Chern-Simons term.
This is a specific feature of the $(2+1)$-dimensional theory.

Consider the Maxwell-Chern-Simons action
\begin{equation}
 S_{\rm M C S} 
= \int {\rm d}^3x 
\(-\fr{1}{4e_{\rm 3D}^2} |f_{\mu\nu}|^2 
+ \fr{k}{4\pi} \epsilon^{\mu\nu\rho} a_{\mu} \der_{\nu} a_{\rho}\)\,.
\end{equation}
The equation of motion of $a_\mu$ is 
\begin{equation}
\fr{1}{e^2_{\rm 3D}}
\der_\nu f^{\nu\mu}
+ \fr{k}{2\pi} \epsilon^{\mu\nu\rho} \der_\nu a_\rho=0\,.
 \label{EOM_3D}
\end{equation}
From an argument similar to the one used in Sec.~\ref{sec:topological}, 
the Chern-Simons term generates the mass of the photon as \cite{Binegar:1981gv,Deser:1982vy,Deser:1981wh}
\begin{equation}
(\der^2 - \Delta_{\rm 3D}^{2}) f_{\mu\nu} = 0, \qquad
\Delta_{\rm 3D}^2 = \left(\fr{k e_{\rm 3D}^2}{2\pi}\right)^{\! 2} \,.
\end{equation}
By using the plane wave ansatz with \er{EOM_3D}, we can show that 
there is one massive degree of freedom in this theory.

In the low-energy limit, the Maxwell-Chern-Simons theory reduces to 
the Chern-Simons theory
\begin{equation}
 S_{\rm CS}
 = \fr{k}{4\pi} \int {\rm d}^3 x \epsilon^{\mu\nu\rho} a_{\mu} \der_{\nu} a_{\rho}\,.
\end{equation}
The observable of the theory is the Wilson loop $W({\cal C}) = {\rm e}^{{\rm i} \int_{\cal C}a_{\mu} {\rm d}x^{\mu} }$.
The correlation function of two Wilson loops has a nonzero fractional phase 
if they are linked to each other,
\begin{equation}
 \vevs{W({\cal C})W({\cal C'})}
 = 
{\rm e}^{\fr{2\pi {\rm i} }{k} \link ({\cal C,C'})} \vevs{W({\cal C'})}
 = 
{\rm e}^{\fr{2\pi {\rm i} }{k} \link ({\cal C,C'})}.
\end{equation}
Therefore, this theory is again in a topologically ordered phase, 
characterized by the type-B SSB of $\bb{Z}_k$ one-form symmetry.
In the present case, in particular, the generator of the 
one-form symmetry is the charged object itself.

\subsubsection{Axion electrodynamics in $(1+1)$ dimensions}
\label{2d}
The final example is an axion-photon system in $(1+1)$ dimensions, 
which we will call the axion electrodynamics in $(1+1)$ dimensions
in this paper.
The higher-form symmetries in this system were studied 
in the context of its fermionic version, the charge-$k$ 
Schwinger model, in Refs.~\cite{Anber:2018jdf,Armoni:2018bga} 
(see also Ref.~\cite{Misumi:2019dwq}).
In $(1+1)$ dimensions, a massless photon has no on-shell dynamical 
degrees of freedom. However, the photon can have a dynamical 
degree of freedom if it becomes massive, e.g., by a topological 
coupling with a scalar field~\cite{Witten:1978bc,Aurilia:1980jz}.

We introduce the following action:
\begin{equation}
 S_{\rm 2D}
 = \int {\rm d}^2x \(
-\fr{1}{4e^2_{\rm 2D}} |f_{\mu\nu}|^2
 - \fr{v_{\rm 2D}^2}{2}|\der_{\mu} \phi|^2 
+ \fr{k}{2\pi}\epsilon^{\mu\nu} a_{\mu} \der_{\nu} \phi
+ \fr{1}{2e_{\rm 2D}^2}\der_\mu (a_\nu f^{\mu\nu})\)\,,
\label{S_2D}
\end{equation}
where $\phi$ is a $2\pi$ periodic scalar field,
which is the $(1+1)$-dimensional version of $(D-2)$-form,
$e_{\rm 2D}$ is a coupling constant, and 
$v_{\rm 2D}$ is some constant. 
The third term $\fr{k}{2\pi}\epsilon^{\mu\nu} a_{\mu} \der_{\nu} \phi$
is a topological coupling between the photon and the scalar field.
The last term is a boundary term for the kinetic term of the photon,
which is needed 
to have a consistent energy momentum tensor with the equation of motion
of $a_{\mu}$~\cite{Brown:1987dd,Brown:1988kg,Duff:1989ah,Duncan:1989ug,Duncan:1990fr,Kaloper:2008fb,Kaloper:2008qs},
but it will be irrelevant to the following discussion.

The equations of motion of $a_{\mu} $ and $\phi$ are
\begin{equation}
  \fr{1}{e_{\rm 2D}^2} \der_{\nu} f^{\nu\mu} 
  + \fr{k}{2\pi} \epsilon^{\mu\nu} \der_{\nu} \phi = 0\,, \qquad
  v_{\rm 2D}^2 \der^{\mu} \der_{\mu} \phi 
  + \fr{k}{2\pi} \epsilon^{\mu\nu} \der_{\mu} a_{\nu} = 0\,,
\label{EOM_2D}
\end{equation}
which can be rewritten as
\begin{equation}
(\der^2
- \Delta_{\rm 2D}^{2}) f_{\mu\nu} = 0, \qquad
(\der^2 - \Delta_{\rm 2D}^{2}) \der_\nu\phi = 0,
\end{equation}
respectively, where
\begin{equation}
\Delta_{\rm 2D}^2 = \left( \fr{k e_{\rm 2D}}{2\pi v_{\rm 2D}} \right)^{\! 2}\,.
\end{equation}
The Bianchi identity with the local solutions of \er{EOM_2D}
after gauge fixing,
\begin{equation}
 f^{\mu\nu} = \fr{e^2_{\rm 2D} k}{2\pi} \epsilon^{\mu\nu}\phi\,,
\qquad
\der^\mu \phi = - \fr{v_{\rm 2D}^2k}{2\pi}\epsilon^{\mu\nu}a_\nu\,,
\label{relation_2D}
\end{equation}
leads to
\begin{equation}
\fr{k}{2\pi} \der^{\nu} a_{\nu}  =0\,. 
\end{equation}
Note that the solution of the equation of motion of $a_{\mu}$ does not 
lead to any constraint on $\phi$.
Therefore, the physical degrees of freedom is one, since $\phi$
is constrained by $ f^{\mu\nu}$ in \er{EOM_2D}.

The low-energy limit is again described by the topological coupling
\begin{equation}
 S_{\rm 2D,top}
 = \fr{k}{2\pi} \int {\rm d}^2 x \epsilon^{\mu\nu} a_{\mu} \der_{\nu} \phi\,.
\end{equation}
The observables are the Wilson loop 
$W({\cal C}) = {\rm e}^{{\rm i}\int_{\cal C} a}$ 
and a two-point object
$I({\cal P,P'}) = {\rm e}^{{\rm i}\phi ({\cal P}) - {\rm i} \phi({\cal P}')}$.
Of course, a one-point object can also be an observable.
However, it is convenient to use this two-point object, 
since it can measure the difference of the values of $\phi$ 
separated by the Wilson loop.

The correlation function is 
\begin{equation}
 \vevs{ W({\cal C})I({\cal P,P'}) } 
=  {\rm e}^{\fr{2\pi {\rm i} }{k} \link({\cal C}, ({\cal P,P'}))},
\end{equation}
where the linking of the loop ${\cal C}$ and the two points $({\cal P,P'})$
denotes the configuration where the point ${\cal P}$ intersects with 
${\cal C}$ when ${\cal P}$ is continuously moved to ${\cal P}'$.
The phase of this system is classified by the type-B SSB of the 
$\bb{Z}_k$ one- and zero-form symmetries~\cite{Pantev:2005rh,Pantev:2005zs,Pantev:2005wj,Gaiotto:2014kfa,Sharpe:2015mja,Hidaka:2019mfm}.

\section{Topological mass generation in gapless systems}
\label{sec:gapless}
From now on, we consider the topological mass generation in gapless systems.
As such an example, we focus on the axion electrodynamics with level $k$
in (3+1) dimensions in a background magnetic field~\cite{Sogabe:2019gif, Brauner:2017uiu} 
or a spatially varying axion field~\cite{Yamamoto:2015maz, Ozaki:2016vwu}.
In these backgrounds, there appear a gapped mode and a gapless mode with the 
quadratic dispersion relation $\omega \sim p^2$ depending on the helicity states. 
It has also been argued that the latter gapless mode may be understood as 
the type-B NG mode associated with the spontaneous breaking  of a 
one-form symmetry \cite{Sogabe:2019gif}. We here argue that the helicity-dependent 
mass generation can be explained from a topological viewpoint. 

The action is given by 
\begin{equation}
 S_{{\rm EM}, \phi} [\phi,a]
=   \int {\rm d}^4 x \(- \fr{1}{4e^2} |f_{\mu\nu}|^2
 - \fr{v^2}{2} |\der_{\mu} \phi|^2
+ \fr{k}{16\pi^2}  \phi f_{\mu\nu}  \tilde{f}^{\mu\nu} \) \,,
\label{axion_elemag}
\end{equation}
where $\phi$ is a $2\pi$ periodic pseudoscalar field and 
$ \tilde{f}^{\mu\nu} \equiv \frac{1}{2} \epsilon^{\mu \nu \rho \sigma} f_{\rho \sigma}$. 
The constant $k$ is again restricted as $k \in \bb{Z}$ owing to the 
invariance under the large gauge transformation of $a_{\mu}$ and 
the periodicity $\phi \to \phi +2\pi$.  
The last term $\fr{k}{16\pi^2} \phi f_{\mu\nu} \tilde{f}^{\mu\nu}$ is the 
topological term that does not depend on the metric of spacetime. 
Unlike the previous examples in gapped systems, this is a cubic interaction 
term and it does not directly contribute to the dispersion relations of 
excitations in the system. 
However, it becomes quadratic in dynamical fields in the presence of a 
background magnetic field (e.g., $\vevs{f_{12}}\neq 0$) 
or a spatially varying axion field (e.g., $\vevs{\der_3 \phi} \neq 0$), 
which can modify their dispersion relations.  
We note in passing that the same type of action appears in QCD coupled to 
QED at low energy, where the role of $\phi$ is played by charge neutral 
pion $\pi_0$ and $|k|=1$.

Without the topological term, this system would have a shift symmetry
$\phi \rightarrow \phi + c$ with $c$ being a constant. 
The Noether current associated with this
symmetry is $j^{\mu}_5 = -2v^2 \der^{\mu} \phi$ that satisfies
$\der_{\mu} j^{\mu}_5 = 0$.  
Here, we chose the normalization of $j_5^{\mu}$ such that, when $|k|=1$, 
it matches that of the axial current carried by pions in the case of QCD.  
However, this shift symmetry is broken by the
presence of the topological term, and consequently, the conservation law
is modified to
\begin{equation}
\label{anomaly}
\der_{\mu} j^{\mu}_5 = \fr{k}{8\pi^2} f_{\mu\nu}  \tilde{f}^{\mu\nu}\,.
\end{equation}
This can be regarded as the chiral anomaly of this system.

In the following, we will clarify the mechanism of the helicity-dependent mass generation 
and the properties of this system in detail.

\subsection{Mass generation and gapless modes}

\subsubsection{Background magnetic field}
First, we consider the system with a homogeneous background magnetic field,
$\vevs{f_{12}} =B_z > 0$.
The equation of motion of $\phi$, $f_{03} = -e_z$ 
and $\tilde{f}_{12} = \epsilon_{1203} f^{03} = - e_z$
at the linearized level read
\begin{equation}
\begin{split}
v^2 \der^{\mu} \der_{\mu} \phi 
- \fr{k B_z}{4\pi^2} e_z =0\,,
\end{split}
\end{equation}
\begin{equation}
\begin{split}
  -\fr{1}{e^2} \der^{\mu}\der_{\mu} e_z 
+ \fr{k B_z}{4\pi^2} (- \der_t^2 +\der_z^2) \phi =0\,.
\end{split}
\end{equation}
The equations of motion in momentum space can be summarized 
in the matrix form
\begin{equation}
\mtx{v^2 (\omega^2 - \bs{p}^2) 
& - \fr{kB_z}{4\pi^2}
\\
-\fr{kB_z}{4\pi^2}(\omega^2 - p_z^2) 
& 
\fr{1}{e^2}(\omega^2 - \bs{p}^2)} 
\mtx{\phi \\ e_z} =0 \,,
\end{equation}
where $p^{\mu}  = (\omega,\bs{p})=(\omega, p_x, p_y, p_z)$.

One can show that there are one gapless mode with the quadratic dispersion 
(in the transverse direction with respect to $B_z$) and one gapped mode,%
\footnote{The longitudinal component of the dispersion relation for the 
gapless mode $\omega_{{\rm gapless}, \parallel}^2 = p_z^2$ is due to the 
anomalous charge $n = -\fr{kB_z}{4\pi^2} \der_z \phi$ in the presence of the 
background magnetic field. When one further introduces a background charge 
such that the local charge neutrality condition is satisfied, then this contribution 
would vanish \cite{Sogabe:2019gif,Brauner:2017uiu}.}
\begin{equation}
\label{dispersion_B}
 \omega^2_{\rm gapless}
 =  p_z^2 + \fr{1}{\Delta_B^2}{\bm p}_{\perp}^4
+{\cal O}(p^6)\,,
\qquad
 \omega^2_{\rm gapped}
 = \Delta_B^2 + 2 {\bm p}_{\perp}^2 + p_z^2
+{\cal O}(p^4)\,,
\end{equation}
where ${\bm p}_{\perp}=(p_x,p_y,0)$ and
\begin{equation}
\Delta_B^2 = \left(\fr{k e B_z}{4\pi^2 v}\right)^{\! 2}\,.
\end{equation}
The corresponding eigenvectors are
\begin{equation}
\mtx{\phi \\ e_z}
\propto
\mtx{\fr{\Delta_B}{ev}+ {\cal O}(p^2)
\\
\fr{{\bm p}_{\perp}^4}{\Delta_B^2}  - {\bm p}_{\perp}^2 
+ {\cal O} (p^6)}\,, 
\qquad
\mtx{\phi \\ e_z}
\propto
\mtx{\fr{\Delta_B}{ev}+ {\cal O}(p^2)
\\
\Delta_B^2 + {\bm p}_{\perp}^2 + {\cal O} (p^4)} \,,
\label{eigenvector_B}
\end{equation} 
for the gapless and gapped modes, respectively.
The gapless mode with the quadratic dispersion in the transverse
direction in Eq.~(\ref{dispersion_B}) can be understood as the type-B NG
mode associated with the spontaneous breaking of the one-form symmetry
\cite{Sogabe:2019gif}.

We now show that the gapped mode in \er{eigenvector_B} is related to the
gapped mode in the axion electrodynamics in $(1+1)$ dimensions in
Sec.~\ref{2d} via dimensional reduction. 
A simple way to see this is that the topological term 
$\epsilon^{\mu\nu 12} \phi f_{\mu\nu}\vevs{f_{12}}$ 
in the present theory in the external magnetic
field has the same structure as the topological term $\epsilon^{\mu\nu}
\phi f_{\mu \nu}$ in the axion electrodynamics in $(1+1)$ dimensions.
To be more concrete, consider the gapped mode in $z$ direction by
setting $p^\mu = (\omega, 0,0,p_z)$. Then, the eigenstate along the $z$
direction, $e_z = \fr{e^2 k B_z}{4\pi^2} \phi$, has the same structure
as \er{relation_2D} but with the additional factor $\frac{B_z}{2\pi}$.
This factor, as well as the ratio between the mass gaps of the two
theories,
\begin{equation}
\fr{\Delta_B}{\Delta_{\rm 2D}} = \fr{B_z}{2\pi}\,,
\end{equation}
can simply be understood as the Landau degeneracy in the transverse direction 
with respect to $B_z$. 

In this way, the mass generation in the axion electrodynamics with the external 
magnetic field in $(3+1)$ dimensions is related to the one in the axion 
electrodynamics in $(1+1)$ dimensions. As photons have only one physical 
degree of freedom in the latter, this argument explains that only one of the two 
helicity states acquires the mass gap in the former, and hence, this is a 
helicity-dependent mass generation.

\subsubsection{Spatially varying axion background}
We can similarly discuss the mass generation of photons 
in the spatially varying axion background field where 
$\vevs{\der_i \phi} \neq 0$.
In this background, the equation of motion of $a_{\mu}$ is
\begin{equation}
\begin{split}
\fr{1}{e^2}\der_{\mu} f^{\mu\nu} -
\fr{k}{4\pi^2} \fr{\epsilon^{i \nu\rho\sigma}}{2}\vevs{ \der_i
 \phi}f_{\rho\sigma}=0\,.
\end{split}
\end{equation}
In the following, we assume that the variation of the axion field is 
positive and homogeneous along the $z$ direction, $\vevs{\der_z \phi} > 0$.
In this case, there is a gapped mode of a linear combination of $e_x$ and $e_y$ 
propagating in the $z$ direction \cite{Yamamoto:2015maz, Ozaki:2016vwu}.

The equations of motion of $e_x$ and $e_y$ in momentum space are summarized as
\begin{equation}
\mtx{
\fr{1}{e^2} (-\omega^2 + p_y^2+ p_z^2) 
&
-\fr{1}{e^2} p_x p_y +\fr{{\rm i} k}{4\pi^2} \vevs{\der_z \phi} \omega
\\
-\fr{1}{e^2} p_x p_y 
-\fr{{\rm i} k}{4\pi^2} \vevs{\der_z \phi} \omega 
&
\fr{1}{e^2} (-\omega^2 + p_x^2+ p_z^2) 
}
\mtx{e_x \\ e_y} =0 \,.
\end{equation}
The dispersion relations can be found as
\begin{equation}
 \omega^2_{\rm gapless} = \fr{1}{\Delta^2_\phi} \bs{p}^2 p_z^2
+{\cal O}(p^6)\,,
\qquad
\omega^2_{\rm gapped}
 = \Delta_\phi^2 + {\bm p}_{\perp}^2 + 2p_z^2+{\cal O}(p^4)\,,
\end{equation}
where
\begin{equation}
\Delta_{\phi}^2 = \left( \fr{k e^2}{4\pi^2} \vev{\der_z \phi} \right)^{\! 2} \,.
\end{equation}
The corresponding eigenvectors are
\begin{equation}
\label{eigenvector_phi}
\mtx{e_x \\ e_y}
\propto
\mtx{p_x p_y -{\rm i} |\bs{p}||p_z|+{\cal O}(p^4) \\ 
p_y^2+ p_z^2 - \fr{1}{\Delta^2_\phi} \bs{p}^2p^2_z +{\cal O}(p^6)}\,, 
\quad
\mtx{e_x \\ e_y}
\propto
\mtx{p_x p_y -{\rm i}
(\Delta_\phi^2 + \tf{{\bm p}_\perp^2 + 2p_z^2}{2})
+{\cal O}(p^4)
 \\ 
-(\Delta_\phi^2+ p_x^2 + p_z^2)+{\cal O}(p^4)}\,,
\end{equation}
for the gapless and gapped modes, respectively. 

This gapped mode is related to the one in the Maxwell-Chern-Simons theory in (2+1) dimensions 
in Sec.~\ref{sec:MCS}. One can see that the topological coupling 
$\fr{k}{4\pi^2}\epsilon^{3\nu\rho\sigma}\vevs{\der_z \phi}$ has the same structure 
as the topological coupling $\fr{k}{2\pi} \epsilon^{\nu\rho\sigma}$ in the 
Maxwell-Chern-Simons theory. For the gapped mode, e.g., along the $y$ direction 
with the momentum $p^\mu = (\omega, 0,p_y,0)$, \er{eigenvector_phi} becomes
\begin{equation}
 \Delta_{\phi} e_x = {\rm i}\omega_{\rm gapped} e_y\,.
\end{equation}
This takes the same form as \er{EOM_3D} with $\mu = y$ and $p_x =0$
up to the factor $- \frac{\vevs{\der_z \phi}}{2\pi}$. This factor is reflected 
in the difference of the gaps between the two theories,
\begin{equation}
\fr{\Delta_{\phi}}{\Delta_{\rm 3D}} = \fr{\vevs{\der_z \phi}}{2\pi}\,.
\end{equation}

As photons have only one physical degree of freedom in the Maxwell-Chern-Simons 
theory in $(2+1)$ dimensions, this argument also explains that only one helicity state 
of photons acquires the mass gap in the system with the spatially varying axion field 
in $(3+1)$ dimensions.

\subsection{Spontaneous breaking of higher-form $\bb{Z}_k$ symmetries}
Here, we discuss higher-form symmetries and their breaking of this theory in the
presence of the background magnetic field or the spatially varying axion field.

\subsubsection{Higher-form symmetries without background fields}

We first review that the theory has $\bb{Z}_k $ zero- and one-form symmetries 
as well as $\U (1)$ one- and two-form symmetries~\cite{Hidaka:2020iaz}. 
For simplicity, we first consider the case without background fields and 
we will discuss the case with background fields later in Sec.~\ref{sec:background}.
By using the equations of motion of $\phi$ and $a_\mu$ derived 
from the action (\ref{axion_elemag}), we can show that there are 
conserved currents of electric zero- and one-form symmetries,
\begin{equation}
 j_{{\rm E} \phi }^\mu = - v^2 \der^\mu \phi - \fr{k}{16\pi^2} 
\epsilon^{\mu\nu\rho\sigma} a_\nu f_{\rho\sigma}\,, \qquad
 j_{{\rm E} a}^{\mu\nu}
= -\fr{1}{e^2} f^{\mu\nu} + \fr{k}{8\pi^2}\epsilon^{\mu\nu\rho\sigma}
\phi f_{\rho\sigma}\,.
\end{equation}
In addition, the Bianchi identities for $\phi$ and $a_\mu$ give the 
following conserved currents of magnetic two- and one-form symmetries:
\begin{equation}
 j_{{\rm M} \phi }^{\mu\nu\rho}
 = \fr{1}{2\pi} \epsilon^{\mu\nu\rho\sigma} \der_\sigma \phi\,, \qquad
 j_{{\rm M} a}^{\mu\nu}
 = \fr{1}{2\pi} \epsilon^{\mu\nu\rho\sigma} \der_\rho a_\sigma\,,
\end{equation}
respectively.
Note that the normalizations of the currents are determined 
by the flux quantization conditions.
We can construct topological objects from these currents,
\begin{align}
 U_{{\rm E} \phi }({\rm e}^{ {\rm i}\alpha_{{\rm E} \phi }}, {\cal V}) 
&= {\rm e}^{ -{\rm i}\alpha_{{\rm E} \phi } \int_{\cal V} 
\fr{\epsilon_{\mu\nu\rho\sigma}}{3!} j_{{\rm E} \phi }^{\mu} 
 {\rm d}S^{\nu \rho \sigma}}, 
\label{210217.2210}
\\
 U_{{\rm E} a}({\rm e}^{ {\rm i}\alpha_{{\rm E} a}}, {\cal S}) 
&= {\rm e}^{-{\rm i}\alpha_{{\rm E} a} \int_{\cal S} 
\fr{\epsilon_{\mu\nu\rho\sigma}}{2! 2!} j_{{\rm E} a}^{\mu\nu} 
 {\rm d}S^{\rho \sigma}},
\\
 U_{{\rm M} \phi }({\rm e}^{ {\rm i}\alpha_{{\rm M} \phi }}, {\cal C}) 
&= {\rm e}^{-{\rm i}\alpha_{{\rm M} \phi } \int_{\cal C} 
\fr{\epsilon_{\mu\nu\rho\sigma}}{3!} j_{{\rm M} \phi }^{\mu\nu\rho} 
{\rm d}x^\sigma},
\\
 U_{{\rm M} a}({\rm e}^{{\rm i}\alpha_{{\rm M} a}}, {\cal S}) 
&= {\rm e}^{- {\rm i}\alpha_{{\rm M} a} \int_{\cal S} 
\fr{\epsilon_{\mu\nu\rho\sigma}}{2! 2!} j_{{\rm M} a}^{\mu\nu} 
 {\rm d}S^{\rho \sigma}},
\label{210217.2211}
\end{align}
where ${\rm e}^{ {\rm i}\alpha_{{\rm E} \phi }}, {\rm e}^{ {\rm i}\alpha_{{\rm E} a}},
{\rm e}^{ {\rm i}\alpha_{{\rm M} \phi}}, {\rm e}^{ {\rm i}\alpha_{{\rm M} a}} \in \U(1)$ 
parametrize the symmetry generators and ${\cal V}$ is a three-dimensional 
closed subspace without boundaries. They are topological since they are 
invariant under small deformations of ${\cal V}$, ${\cal S}$, and ${\cal C}$
by the equations of motion or Bianchi identities.
We put the minus signs in the prefactors of \ers{210217.2210}--\eqref{210217.2211}
for convenience.

However, $ U_{{\rm E} \phi }$ and $ U_{{\rm E} a}$ are not physical 
observables for generic $\alpha_{{\rm E} \phi }$ and $\alpha_{{\rm E} a}$ 
due to the large gauge transformation of $a_{\mu}$ and the $2\pi$ 
periodicity of $\phi$. The invariance under the large gauge transformation
and the periodicity requires that the parameters 
${\rm e}^{ {\rm i}\alpha_{{\rm E} \phi }}$
and ${\rm e}^{ {\rm i}\alpha_{{\rm E} a}}$ are 
constrained as ${\rm e}^{ {\rm i}\alpha_{{\rm E} \phi }}, {\rm e}^{ {\rm i}\alpha_{{\rm E} a}} \in \bb{Z}_k$.
Note that this argument is similar to the quantization of 
the Chern-Simons level in $(2+1)$ dimensions~\cite{Dijkgraaf:1989pz}.
Therefore, the symmetry groups of $ U_{{\rm E} \phi }$ and $ U_{{\rm E} a}$
are $\bb{Z}_k$, while the symmetry groups of 
$U_{{\rm M} \phi }$ and $ U_{{\rm M} a}$ are $\U(1)$.

Let us explain the detail.
For $U_{{\rm E} \phi }$, the relevant term is 
${\rm e}^{-{\rm i}k\fr{\alpha_{{\rm E} \phi }}{8\pi^2} \int_{\cal V} 
 a_\nu \der_\rho a_\sigma {\rm d}S^{\nu \rho \sigma}} $,
which is subject to the large gauge transformation.
In order to make the integrand manifestly gauge invariant, 
we introduce a four-dimensional space $X_4$
whose boundary is ${\cal V}$, 
$\der X_4 = {\cal V}$, to write
\begin{equation}
 {\rm e}^{- {\rm i}k\fr{\alpha_{{\rm E} \phi }}{8\pi^2} \int_{\cal V} 
a_\nu \der_\rho a_\sigma {\rm d}S^{\nu \rho \sigma}} 
=
 {\rm e}^{ - {\rm i}k\fr{\alpha_{{\rm E} \phi }}{16\pi^2}
 \int_{X_4} {\rm d}^4 x
f_{\mu \nu} \tilde f^{\mu \nu}}.
\end{equation}
We require that the term should not depend on the choice 
of the four-dimensional space.
This requirement leads to the constraint on the parameter 
${\rm e}^{{\rm i}\alpha_{{\rm E} \phi }}$.
By choosing a four-dimensional space $X_4'$
different from $X_4$, such that $\der X_4' = {\cal V}$,
we have
\begin{equation}
 {\rm e}^{- {\rm i}k\fr{\alpha_{{\rm E} \phi }}{8\pi^2} \int_{\cal V} 
 a_\nu \der_\rho a_\sigma {\rm d}S^{\nu \rho \sigma}} 
=
 {\rm e}^{- {\rm i}k\fr{\alpha_{{\rm E} \phi }}{16\pi^2} \int_{X_4} 
{\rm d}^4 x f_{\mu \nu} \tilde f^{\mu \nu}}
=
 {\rm e}^{- {\rm i}k\fr{\alpha_{{\rm E} \phi }}{16\pi^2} \int_{X_4'} 
{\rm d}^4 x f_{\mu \nu} \tilde f^{\mu \nu}}.
\end{equation}
Therefore, the integral is independent of the choice of 
the redundant space if the following condition is satisfied:
\begin{equation}
 {\rm e}^{- {\rm i}k\fr{\alpha_{{\rm E} \phi }}{16\pi^2} \int_{Y_4} 
{\rm d}^4 x f_{\mu \nu} \tilde f^{\mu \nu} }= 1,
\label{condition}
\end{equation}
where $Y_4= X_4 - X_4'$ is a four-dimensional manifold 
that has no boundary, $\der Y_4 = \der X_4 - \der X_4' =0$,
but it can generally have cycles which can be wrapped by the field 
strength $f_{\mu\nu}$.
Since $ \int_{Y_4} {\rm d}^4 x f_{\mu \nu} \tilde f^{\mu \nu} \in 16 \pi^2 \bb{Z}$ 
by the flux quantization condition, the parameter $\alpha_{{\rm E} \phi }$ 
is constrained as ${\rm e}^{ {\rm i}\alpha_{{\rm E} \phi }} \in \bb{Z}_k$
in order to satisfy the condition (\ref{condition}).

For $U_{{\rm E} a}$, the relevant term is 
${\rm e}^{\fr{{\rm i}k\alpha_{{\rm E} a}}{8\pi^2} 
\int_{\cal S} \phi f_{\mu\nu} {\rm d}S^{\mu\nu}}$, 
which changes under the transformation $\phi \to \phi +2\pi$ as
\begin{equation}
 {\rm e}^{\fr{ {\rm i}k\alpha_{{\rm E} a}}{8\pi^2}
\int_{\cal S} \phi f_{\mu\nu} {\rm d}S^{\mu\nu}}
 \to 
{\rm e}^{\fr{ {\rm i}k\alpha_{{\rm E} a}}{4\pi}
\int_{\cal S} f_{\mu\nu} {\rm d}S^{\mu\nu}}
{\rm e}^{\fr{ {\rm i}k\alpha_{{\rm E} a}}{8\pi^2}
\int_{\cal S} \phi f_{\mu\nu} {\rm d}S^{\mu\nu}}.
\end{equation}
To make it invariant under this transformation, we require 
\begin{equation}
 {\rm e}^{\fr{{\rm i}k \alpha_{{\rm E} a}}{4\pi}
\int_{\cal S} f_{\mu\nu} {\rm d}S^{\mu\nu}} = 1,
\end{equation}
which constrains the parameter $\alpha_{{\rm E} a}$ 
as ${\rm e}^{ {\rm i}\alpha_{{\rm E} a}} \in \bb{Z}_k$ by 
the flux quantization condition $ 
\fr{1}{2!}\int_{\cal S} f_{\mu\nu} {\rm d}S^{\mu\nu} \in 2 \pi \bb{Z}$ .

\subsubsection{Symmetry transformations}
\label{sec:transformation}

Next, we consider the charged objects that transform under 
the extended topological objects:
\begin{itemize}
\item 
For the electric $\bb{Z}_k$ zero-form symmetry generated by $U_{{\rm E} \phi }$,
the charged object is a particle at a point ${\cal P}$ in the spacetime,
\begin{equation}
 I(q_{{\rm E} \phi }, {\cal P}) = {\rm e}^{ {\rm i}q_{{\rm E} \phi } \phi ({\cal P})},
\end{equation}
where $q_{{\rm E} \phi } \in \bb{Z}$ is the charge of the particle.
The transformation law is given by the correlation function
\begin{equation}
 \vevs{U_{{\rm E} \phi } 
({\rm e}^{ {\rm i}\alpha_{{\rm E} \phi }},{\cal V})I(q_{{\rm E} \phi }, {\cal P}) }
= {\rm e}^{{\rm i}\alpha_{{\rm E} \phi }q_{{\rm E} \phi }\link({\cal V,P})}
\vevs{I(q_{{\rm E} \phi }, {\cal P}) }.
\label{correlation_I}
\end{equation}
The derivation of this relation will be given at the end of 
Sec.~\ref{sec:transformation}. 
Since ${\rm e}^{ {\rm i}\alpha_{{\rm E} \phi }} \in \bb{Z}_k$ 
and $q_{{\rm E} \phi }\link({\cal V,P}) \in \bb{Z}$, we have 
${\rm e}^{{\rm i}\alpha_{{\rm E} \phi }q_{{\rm E} \phi }\link({\cal V,P})} \in \bb{Z}_k$.

\item 
For the electric $\bb{Z}_k$ one-form symmetry generated by $U_{{\rm E} a}$, 
the charged object is the Wilson loop
$W(q_{{\rm E} a}, {\cal C}) = {\rm e}^{{\rm i} q_{{\rm E} a} \int_{\cal C} a_\mu dx^{\mu }}$.
The symmetry transformation is given by
\begin{equation}
 \vevs{U_{{\rm E} a} ({\rm e}^{ {\rm i} \alpha_{{\rm E} a}}, {\cal S}) 
W(q_{{\rm E} a}, {\cal C})}
 = 
{\rm e}^{ {\rm i} \alpha_{{\rm E} a}q_{{\rm E} a} \link( {\cal S,C})}
\vevs{
W(q_{{\rm E} a}, {\cal C})} .
\label{correlation_W2}
\end{equation}

\item 
For the magnetic $\U(1)$ two-form symmetry generated by $U_{{\rm M} \phi}$,
the charged object is a vortex world surface $V(q_{{\rm M} \phi }, {\cal S})$,
which can be characterized by the winding number around a loop ${\cal C}$,
\begin{equation}
\int_{\cal C} \der_\mu \phi {\rm d}x^\mu  = 2\pi q_{{\rm M} \phi} 
\link ({\cal C,S})\,.
\end{equation}
The symmetry transformation is
\begin{equation}
 \vevs{U_{{\rm M} \phi} ({\rm e}^{{\rm i}\alpha_{{\rm M} \phi } }, {\cal C}) 
V(q_{{\rm M} \phi }, {\cal S})} 
= 
{\rm e}^{{\rm i}\alpha_{{\rm M} \phi } q_{{\rm M} \phi } \link({\cal C,S})} 
\vevs{V(q_{{\rm M} \phi }, {\cal S})} .
\end{equation}

\item 
For the magnetic $\U(1)$ one-form symmetry generated by $U_{{\rm M} a}$,
the charged object is the 't Hooft loop, i.e., a monopole world line 
$T(q_{{\rm M} a}, {\cal C})$, which can be characterized by 
the presence of the magnetic flux through a closed surface ${\cal S}$
(e.g, a sphere $S^2$),
\begin{equation}
\fr{1}{2}\int_{\cal S} f_{\mu\nu}(x) {\rm d}S^{\mu\nu}
= 2\pi q_{{\rm M} a} \link ({\cal S,C})\,.
\end{equation}
The symmetry transformation is 
\begin{equation}
 \vevs{U_{{\rm M} a} ({\rm e}^{{\rm i}\alpha_{{\rm M} a} }, {\cal C}) 
T(q_{{\rm M} a}, {\cal C})} 
= 
{\rm e}^{{\rm i}\alpha_{{\rm M} a} q_{{\rm M} a} \link({\cal S,C})} 
\vevs{T(q_{{\rm M} a}, {\cal C})} .
\end{equation}
\end{itemize}

Let us derive Eq.~(\ref{correlation_I}) as an example. 
The key relation is the following: 
\begin{equation}
S_{{\rm EM},\phi} [\phi,a]
-  
\alpha_{{\rm E}\phi }
\int  {\rm d}^4x 
j^{\mu}_{{\rm E} \phi}
J_\mu ({\cal V})
= 
S_{{\rm EM},\phi} 
[\phi - \alpha_{{\rm E}\phi} J(\Omega_{\cal V}),a]
\label{key}
\end{equation}
up to the trivial divergence that can be regularized 
by adding a local counterterm.
Here, $\Omega_{\cal V}$ is a four-dimensional subspace 
whose boundary is ${\cal V}$, 
$\der \Omega_{\cal V} = {\cal V}$, and the symbols
$J_{\mu}({\cal V})$ and $J(\Omega_{\cal V})$ are 
abbreviations of 
$J_{\mu}(x; {\cal V})$ and $J(x;\Omega_{\cal V})$, 
which are defined by
\begin{equation}
 J_\mu (x; {\cal V})
 = \fr{\epsilon_{\mu\nu\rho\sigma}}{3!}
\int_{{\cal V}}\delta^4 (x-y) {\rm d} S^{\nu\rho\sigma}(y)
\end{equation}
and
\begin{equation}
 J(x; \Omega_{\cal V})
 =  
\int_{\Omega_{\cal V}}{\rm d}^4y \delta^4 (x-y) \,,
\end{equation}
respectively. 
As in the case of $J^{\mu\nu} ({\cal S_C})$ in \er{delta_current},
they are introduced such that 
\begin{equation}
 \fr{1}{3!}\int_{\cal V} 
\epsilon_{\mu\nu\rho\sigma} j^\mu_{{\rm E} \phi} {\rm d} S^{\nu\rho\sigma}
= \int {\rm d}^4 x j^\mu_{{\rm E} \phi} J_\mu ({\cal V})\,, 
\qquad
\int_{\Omega_{\cal V}} {\rm d}^4 x
\der_{\mu} j^\mu_{{\rm E} \phi}
=  \int {\rm d}^4 x
J(\Omega_{\cal V})\der_{\mu} j^\mu_{{\rm E} \phi} \,. 
\end{equation}
In particular, 
$J(\Omega_{\cal V})$ satisfies
\begin{equation}
 \der_{\mu}  J(\Omega_{\cal V})  = J_\mu ({\cal V}). 
\end{equation}

Now, the left-hand side of \er{correlation_I} can be evaluated by the redefinition 
$\phi(x) -\alpha_{{\rm E}\phi } J(\Omega_{\cal V}) \to \phi(x)$ on the right-hand side
of Eq.~(\ref{key}) as 
\begin{equation}
 \vevs{U_{{\rm E} \phi } 
({\rm e}^{ {\rm i}\alpha_{{\rm E} \phi }},{\cal V})
I(q_{{\rm E} \phi }, {\cal P}) }
= 
{\rm e}^{ {\rm i} \alpha_{{\rm E} \phi } q_{{\rm E} \phi} 
J({\cal P}; \Omega_{\cal V})} 
\vevs{I(q_{{\rm E} \phi }, {\cal P}) }.
\end{equation}
Since $J({\cal P}; \Omega_{\cal V}) \in \bb{Z}$ 
is the intersection number of ${\cal P}$ and $\Omega_{\cal V}$,
it counts the linking number $\link ({\cal V,P})$.
Therefore, we obtain the relation (\ref{correlation_I}).

\subsubsection{Higher-form symmetries with background fields}
\label{sec:background}
We can describe the external magnetic field and spatially varying 
axion field in terms of background gauge fields coupled to the 
symmetry generators of the electric zero- and one-form symmetries.
For this purpose, we introduce one- and two-form gauge fields 
${\cal A}_\mu $ and ${\cal B}_{\mu\nu}$.
At the linear order of ${\cal A}_\mu$ and ${\cal B}_{\mu\nu}$,
we can gauge the action (\ref{axion_elemag}) by adding the 
coupling terms $\int  {\rm d}^4 x j_{{\rm E} \phi }^\mu {\cal A}_\mu$
and $\fr{1}{2} \int  {\rm d}^4 x j^{\mu\nu}_{{\rm E} a} {\cal B}_{\mu\nu}$.
However, the invariance under the large gauge transformation of $a_\mu$ 
and the periodicity $\phi \to \phi + 2\pi$ requires that 
the background gauge fields should be flat connections satisfying 
\begin{equation}
k {\cal A}_\mu = \der_\mu \bar {\cal A}
\qtq{and}
k {\cal B}_{\mu\nu} 
= \der_\mu \bar {\cal B}_{\nu} - \der_\nu \bar {\cal B}_{\mu},
\end{equation}
respectively.
Here, $\bar {\cal A}$ and $\bar {\cal B}_{\mu}$ are zero- and one-form gauge fields 
that are normalized by the flux quantization conditions,
$\int_{\cal C} \der_\mu \bar {\cal A} {\rm d}x^{\mu} \in 2\pi \bb{Z}$
and 
$\int_{\cal S} \der_\mu \bar {\cal B}_\nu {\rm d} S^{\mu\nu} \in 2\pi \bb{Z}$.
The couplings of the currents to the background gauge fields 
are invariant under the gauge transformations
\begin{equation}
{\cal A}_\mu \to  {\cal A}_\mu + \der_\mu \Lambda,
\qquad
\bar {\cal A} \to \bar {\cal A} + k \Lambda,
\label{trans_A}
\end{equation}
and
\begin{equation}
{\cal B}_{\mu \nu} \to {\cal B}_{\mu \nu} 
+ \der_\mu \Lambda_\nu - \der_\nu \Lambda_\mu,
\qquad
\bar {\cal B}_\mu  \to \bar {\cal B}_\mu + k \Lambda_\mu .
\label{trans_B}
\end{equation}
Here, $\Lambda$ and $\Lambda_\mu $ are zero- and one-form gauge parameters 
normalized as 
$\int_{\cal C} \der_\mu \Lambda {\rm d}x^{\mu} \in 2\pi \bb{Z}$
and 
$\int_{\cal S} \der_\mu \Lambda_\nu {\rm d} S^{\mu\nu} \in 2\pi \bb{Z}$,
respectively.

Now, let us specify the background gauge field that describes 
the spatially varying axion field or external magnetic field.
For the spatially varying axion field, 
we choose ${\cal A}_\mu =  \delta_\mu^z  \vevs{\der_z \phi (z)} $.
This choice corresponds to $\bar {\cal A} = k \vev{\phi (z)} $ 
satisfying $\int_{\cal C} \der_\mu \bar {\cal A} {\rm d}x^\mu =0$.
The coupling term $\int {\rm d}^4 x j_{{\rm E} \phi }^\mu {\cal A}_\mu$ leads to 
the desired coupling
\begin{equation}
\fr{k}{8 \pi^2}\int {\rm d}^4 x 
\epsilon^{3 \mu \nu \rho} \vevs{\der_z \phi (z)} 
a_\mu \der_\nu a_\rho \,. 
\end{equation}
Similarly, the external magnetic field can be realized by choosing 
${\cal B}_{\mu\nu} =  (\delta_\mu^x \delta_\nu^y - \delta_\mu^y \delta_\nu^x) B_z$ 
with 
$\bar {\cal B}_{\nu} = \tf{1}{2} k (x \delta_\nu^y - y \delta_\nu^x)  B_z$
satisfying 
$\int_{\cal S} \der_\mu \bar {\cal B}_{\nu} {\rm d} S^{\mu\nu} =0 $.
The coupling term $\fr{1}{2}\int {\rm d}^4 x j^{\mu\nu}_{{\rm E} a} {\cal B}_{\mu\nu}$ yields
\begin{equation}
 \fr{k}{4\pi^2} \int 
{\rm d}^4 x \phi f_{03} B_z \,.
\end{equation}

We can also couple the system to ${\cal A}_\mu$ and ${\cal B}_{\mu\nu}$ 
at the nonlinear order. These couplings can be added by the replacements 
$\der_\mu \phi \to \der_\mu \phi - {\cal A}_\mu $ 
and $f_{\mu\nu} \to f_{\mu\nu} - {\cal B}_{\mu\nu}$
in \er{axion_elemag}, respectively.
The dynamical fields transform under the gauge transformations in \ers{trans_A} 
and 
\eqref{trans_B} as 
\begin{equation}
 \phi \to \phi + \Lambda,
\label{trans_phi}
\end{equation}
and 
\begin{equation}
a_\mu \to a_\mu + \Lambda_\mu, 
\label{trans_a}
\end{equation}
respectively.%
\footnote{Technically, the nonlinear gauging of the electric one-form symmetry 
violates the periodicity $\phi \to \phi +2\pi$, and we should gauge 
the two-form symmetry simultaneously~\cite{Hidaka:2020izy}.
However, such a violation is absent if we focus on the external magnetic field 
or the spatially varying axion field discussed in this paper.}
The spatially varying axion field and 
external magnetic field can also be realized by the above choices.

In the gauged action, the magnetic one- and two-form symmetries can be 
explicitly broken in general. The gauge invariance under 
 \ers{trans_A}, 
\eqref{trans_B},
\eqref{trans_phi},
and \eqref{trans_a}
requires the symmetry generators should be 
deformed as 
${\rm e}^{\fr{{\rm i} \alpha_{{\rm M} a} }{4\pi}\int_{\cal S} (f_{\mu\nu} - {\cal B}_{\mu\nu}) {\rm d}S^{\mu\nu}}$
for the magnetic one-form symmetry and 
${\rm e}^{\fr{{\rm i} \alpha_{{\rm M} \phi} }{2\pi}\int_{\cal C}
 (\der_{\mu}\phi  - {\cal A}_{\mu}) {\rm d}x^{\mu}}$
for the magnetic two-form symmetry.
However, the presence of the background fields may violate the $2\pi$ 
periodicity of the parameters $\alpha_{{\rm M} a}$ and $\alpha_{{\rm M} \phi }$ 
by fractional phases due to the relations 
$\fr{1}{2}\int_{\cal S} {\cal B}_{\mu\nu} {\rm d}S^{\mu\nu} \in \fr{2\pi}{k} \bb{Z}$
and 
$\int_{\cal C} {\cal A}_{\mu} {\rm d}x^{\mu} \in\fr{2\pi}{k} \bb{Z}$.
In the case of the spatially varying axion field or the external magnetic field, 
these symmetries are preserved since the background fields do not lead to 
fractional phases.

\subsubsection{Higher-form symmetry breaking}
Here, we explain the higher-form symmetry breaking in the presence 
of the external fields. In this system, the $\bb{Z}_k$ zero- and one-form 
symmetries and $\U(1)$ one-form symmetry can be spontaneously broken.%
\footnote{In relativistic systems, the $\U(1)$ two-form symmetry cannot 
be broken spontaneously in $(3+1)$ dimensions due to the higher-form 
generalization~\cite{Gaiotto:2014kfa,Lake:2018dqm} of the 
Coleman-Mermin-Wagner-Hohenberg theorem~\cite{Coleman:1973ci,Mermin:1966fe,Hohenberg:1967zz}.
In nonrelativistic systems, this theorem may be relaxed due to the 
modification of the propagator of a NG mode~\cite{Watanabe:2019xul}.
It is an open question whether the Coleman-Mermin-Wagner-Hohenberg theorem 
for higher-form symmetries can also be relaxed in nonrelativistic systems.}

We are interested in the breaking of the electric one-form symmetry 
characterized by a nonzero VEV of the Wilson loop.
Consider a pair of static charges with opposite signs.
We assume that the distance of the charges is $R$ 
and they are created at $t = 0$ and annihilated at $t = T$.
Under this assumption, the VEV of the Wilson loop is
\begin{equation}
\langle W(q_{{\rm E} a}, {\cal C}) \rangle
 = {\rm e}^{-TV(R)},
\end{equation}
where $V(R)$ is the potential energy.
To evaluate $V(R)$, we consider the propagator of the 
gapless modes in the static limit $\omega = 0$.

Let us first consider the case with the background magnetic field.
To see whether the symmetry breaking occurs, we look at the kinetic term 
of $\phi$ and $a_\mu$, 
\begin{equation}
\label{L_kin}
{\cal L}_{\rm kin} = \frac{1}{2}\mtx{v \phi & \tf{1}{e} a^\mu}{\cal D}
 \mtx{v \phi \\ \tf{1}{e} a^\mu}\,.
\end{equation}
Here, the matrix ${\cal D}$ in momentum space, 
which corresponds to the inverse propagator, is given by \begin{equation} 
 {\cal D}(p)
 =
 \mtx{-p^2 & 
{\rm i} \tf{\Delta_B}{2 B_z} 
\epsilon_{\nu\rho\sigma\tau}\vevs{f^{\sigma\tau}}p^\rho
\\
-{\rm i} \tf{\Delta_B}{2 B_z} 
\epsilon_{\mu\rho\sigma\tau}\vevs{f^{\sigma\tau}}p^\rho
 & 
-\eta_{\mu\nu}p^2}
\equiv
 \mtx{-p^2 & \bs{b}^\dg
\\
\bs{b} & 
-\eta_{\mu\nu}p^2},
\end{equation}
where 
$\bs{b} = ({\rm i} \Delta_B p^z, 0, 0 ,  -{\rm i} \Delta_B \omega)^T$
and we used the Feynman gauge. 
As we are interested in the static Wilson loop in the infrared (IR) regime,
we consider the propagator at $\omega=0$,
\begin{equation}
 {\cal D}^{-1}({\bm p})
= 
\fr{1}{\bs{p}^4 + \Delta_B^2 p_z^2}
\mtx{
-\bs{p}^2 
& 
-{\rm i} \Delta_B p_z 
&
\bs{0}^T
\\
{\rm i} \Delta_B p_z
&
\bs{p}^2 
&
\bs{0}^T
\\
\bs{0}
&
\bs{0}
&
-\delta_{ij} \fr{\bs{p}^4 + \Delta_B^2 p_z^2}{\bs{p}^2}
}
\,.
\end{equation}
The nontrivial eigenvalues of the matrix due to $B_z$ are
$\pm (\bs{p}^4 + \Delta_B^2 p_z^2)^{-1/2}$. Among them 
the eigenvalue $(\bs{p}^4 + \Delta_B^2 p_z^2)^{-1/2}$
corresponds to an unphysical mode and we discard it.%
\footnote{In order to understand the fact that the eigenstate associated with 
the eigenvalue $(\bs{p}^4 + \Delta_B^2 p_z^2)^{-1/2}$ is unphysical, 
it is convenient to consider the case without the external magnetic field, 
where ${\cal D} = \diag (-p^2, p^2, -p^2,-p^2,-p^2)$.
In this case, one of the eigenvalues of ${\cal D}^{-1}$, 
$-\fr{1}{\omega^2- \bs{p}^2}$, which reduces to $\fr{1}{\bs{p}^2}$ when 
$\omega = 0$, corresponds to an unphysical ghost mode due to 
the gauge-fixing term. In our system, the eigenvalue 
$(\bs{p}^4 + \Delta_B^2 p_z^2)^{-1/2}$ corresponds to this unphysical 
mode, since it becomes $\fr{1}{\bs{p}^2}$ when ${\bm B} = {\bm 0}$.}

The potential $V({\bm x})$ due to this mode in the IR limit is given 
by the inverse Fourier transformation of the propagator at $\omega = 0$,
\begin{equation}
\lim_{|{\bm x}| \rightarrow \infty} V({\bm x}) 
\propto \lim_{|{\bm x}| \rightarrow \infty} \int \frac{{\rm d}^3 {\bm p}}{(2\pi)^3} \frac{1}{(\bs{p}^4 + \Delta_B^2 p_z^2)^{1/2}} {\rm e}^{{\rm i}{\bm p} \cdot {\bm x}}
=0\,,
\end{equation}
and hence, the VEV of the Wilson loop is
\begin{equation}
\langle W \rangle  \sim  \lim_{|{\bm x}| \rightarrow \infty} {\rm e}^{-TV({\bm x})} \neq 0\,.
\end{equation}
Therefore, the electric $\bb{Z}_k$ one-form symmetry is broken spontaneously.

We can also show the spontaneous breaking of the electric one-form symmetry 
in the case with the spatially varying axion background field.
In this case, the matrix in \er{L_kin} can be written as
\begin{equation}
{\cal D}(p)
 = 
\mtx{
-p^2 & 0
\\
0 
& 
-\eta_{\mu\nu} p^2 
-{\rm i} \Delta_\phi \epsilon^{3\mu\rho \nu} p_\rho
}\,.
\end{equation}
In the static limit $\omega = 0$, the nontrivial eigenvalues of 
the propagator due to $\vevs{\der_z \phi}$ is 
$-(\bs{p}^4+\Delta_\phi^2 {\bm p}_{\perp}^2)^{-1/2}$.
Here we discarded the eigenvalue $(\bs{p}^4+\Delta_\phi^2 {\bm p}_{\perp}^2)^{-1/2}$ 
that corresponds to an unphysical mode. 
Similar to the argument above, the potential due to this mode in the limit 
$|{\bm x}| \rightarrow \infty$ vanishes, and thus, the electric $\bb{Z}_k$ one-form 
symmetry is broken spontaneously.

\subsection{Low-energy effective theory and chiral anomaly matching}
Since the system has gapped modes with the mass gap $\Delta_{B, \phi}$, 
one can derive a low-energy effective theory for gapless modes well below 
$\Delta_{B, \phi}$ by integrating out these gapped modes. One expects that 
the chiral anomaly in Eq.~(\ref{anomaly}) should also be encoded in such a 
low-energy effective theory, since it is invariant under the renormalization. 
This is the 't Hooft anomaly matching. In the case of the spatially varying 
axion field, this is satisfied simply by $\phi$ with the linear dispersion relation. 
On the other hand, in the case of the background magnetic field, the chiral 
anomaly matching is satisfied by the gapless mode with the quadratic 
dispersion relation (\ref{dispersion_B}), as we will show below.

Let us consider the case with the homogeneous background magnetic field.
We define complex scalar fields%
\footnote{This procedure is similar to the one used to derive the nonrelativistic limit of 
the relativistic field theory for real scalar fields; see, e.g., Refs.~\cite{Guth:2014hsa, Namjoo:2017nia, Braaten:2018lmj}.}
\begin{equation}
\label{psi}
\psi \equiv \sqrt{\frac{\Delta_B}{2}} {\rm e}^{{\rm i}\Delta_B t} \left(v \phi + {\rm i} \frac{a_z}{e} \right)\,, \qquad
\psi^* \equiv \sqrt{\frac{\Delta_B}{2}} {\rm e}^{-{\rm i}\Delta_B t} \left(v \phi - {\rm i} \frac{a_z}{e} \right)\,,
\end{equation}
from which we can write
\begin{equation}
\label{phi}
v \phi = \frac{1}{\sqrt{2 \Delta_B}}({\rm e}^{-{\rm i}\Delta_B t} \psi + {\rm e}^{{\rm i}\Delta_B t} \psi^*)\,, \qquad
\frac{a_z}{e} = -\frac{\rm i}{\sqrt{2 \Delta_B}}({\rm e}^{-{\rm i}\Delta_B t} \psi - {\rm e}^{{\rm i}\Delta_B t} \psi^*)\,.
\end{equation}
Inserting these relations into the original Lagrangian (\ref{axion_elemag}) 
and retaining the terms at the leading order in derivatives under the counting $\der_t \sim \der_z \sim {\bm \nabla}_{\! \perp}^2$, 
where ${\bm \nabla}_{\! \perp} \equiv (\der_x, \der_y, 0)$, we derive the low-energy effective theory
to the second order in fields,
\begin{align}
{\cal L}_{\rm EFT} &= \frac{\rm i}{2} \psi^* \overleftrightarrow{\der_t} \psi - \frac{1}{\Delta_B} {\bm \nabla}_{\! \perp} \psi^* \cdot {\bm \nabla}_{\! \perp} \psi 
+ \frac{\sqrt{2\Delta_B}}{e} \der_z a^t {\rm Re}\left({\rm e}^{-{\rm i}\Delta_B t} \psi \right)  \nonumber \\
& \quad 
+ \frac{1}{e} \sqrt{\frac{2}{\Delta_B}} \der_z a^{\mu}_{\perp} \der_{\mu} {\rm Im}\left({\rm e}^{-{\rm i}\Delta_B t} \psi \right)  
- \frac{1}{4e^2}|f_{\mu \nu}^{\perp}|^2 \,,
\label{EFT}
\end{align}
up to total derivative terms, where 
$\overleftrightarrow{\der_t} \equiv \overrightarrow{\der_t} - \overleftarrow{\der_t}$,
$V^{\mu}_{\perp} \equiv (V^t, V^x, V^y, 0)$ for a generic vector $V^{\mu}$, 
and $f_{\mu \nu}^{\perp} \equiv \der_{\mu} a_{\nu}^{\perp} - \der_{\nu} a_{\mu}^{\perp}$.
It is clear that we have only gapless modes in this low-energy effective theory. 
In particular, the gapless mode $\psi$ has the quadratic dispersion relation 
(\ref{dispersion_B}) in the transverse direction as it should.

We note that the number of physical degrees of freedom in this effective
theory decreases by one (which corresponds to the gapped mode) compared
with that in the original theory (\ref{axion_elemag}).
This can be understood from the fact that the number of physical degrees
of freedom is equal to the number of real fields when the kinetic term
is second order in time derivatives, while it is equal to the number of
complex fields (or half of the number of real fields) when the kinetic
term is first order in time derivatives \cite{Braaten:2018lmj}.

The symmetries of this effective theory 
that we are interested in are as follows:%
\footnote{We will not discuss the higher-form symmetries of the theory
(\ref{EFT}) because, as we only focus on the kinetic terms, the
higher-form symmetries turn out to be enhanced compared to the full
theory with the interaction terms.
A systematic construction of the full effective field theory based on
the symmetries is deferred to future work.}
\begin{itemize}
\item
Gauge symmetry
\begin{equation}
a_{\mu}^{\perp} \rightarrow a_{\mu}^{\perp} + \der_{\mu}^{\perp}
\lambda\,, 
\qquad 
{\rm Im}\left({\rm e}^{-{\rm i}\Delta_B t} \psi
\right) 
\rightarrow 
{\rm Im}\left({\rm e}^{-{\rm i}\Delta_B t} \psi
\right) + \frac{1}{e}\sqrt{\frac{\Delta_B}{2}} \der_z \lambda\,,
\end{equation}
with $\lambda$ a zero-form gauge parameter, which originates from the gauge
symmetry $a_{\mu} \rightarrow a_{\mu} + \der_{\mu} \lambda$ in the
original theory (\ref{axion_elemag}).

\item
Anomalous shift symmetry
\begin{equation}
{\rm Re}\left({\rm e}^{-{\rm i}\Delta_B t} \psi \right) 
\rightarrow 
{\rm Re}\left({\rm e}^{-{\rm i}\Delta_B t} \psi \right) 
+ \sqrt{\frac{\Delta_B}{2}} v c\,,
\end{equation}
with $c$ a constant, which originates from the anomalous shift symmetry
$\phi \rightarrow \phi + c$ in the original theory
(\ref{axion_elemag}).
\end{itemize}

In this effective theory, the canonical momentum of $\psi$ is 
\begin{equation}
\pi \equiv \frac{\der {\cal L}_{\rm EFT}}{\der (\der_t \psi)}
= 
\frac{\rm i}{2} \psi^* \,.
\end{equation}
From the canonical commutation relation 
$[\psi({\bm x}), \pi({\bm y})] = \frac{{\rm i}}{2} \delta({\bm x} - {\bm y})$,%
\footnote{The factor $1/2$ here is due to the two-sided derivative $\overleftrightarrow{\der_t}$ in Eq.~(\ref{EFT}).}
we find
\begin{equation}
\label{comm}
[\psi({\bm x}), \psi^*({\bm y})] = \delta({\bm x} - {\bm y})\,.
\end{equation}

Alternatively, one can derive the commutation relation (\ref{comm}) by
substituting Eq.~(\ref{phi}) into the commutation relation in the
original theory (\ref{axion_elemag}),
\begin{equation}
v^2 [\phi({\bm x}), \der_t \phi({\bm y})] = {\rm i} \delta({\bm x} - {\bm y})\,,
\end{equation}
and using the relation $v \der_t \phi \approx \Delta_B \frac{a_z}{e}$ at
low energy.

One way to check how the chiral anomaly is satisfied in this effective
theory is to look at the anomalous commutator between the vector charge
$n$ and axial charge $n_5$.
For this purpose, let us derive the expressions of $n$ and $n_5$ in
terms of $\psi$.
The vector charge of the effective theory is given by
\begin{equation}
n \equiv - \frac{\delta S_{\rm EFT}}{\delta a_t} 
= - \frac{\sqrt{2 \Delta_B}}{e} \der_z \left[{\rm Re}\left({\rm e}^{-{\rm i}\Delta_B t} \psi \right) \right] \,.
\end{equation}
On the other hand, the axial charge in the theory (\ref{axion_elemag}) is $n_5 = 2 v^2 \der_t \phi$, 
which, in terms of $\psi$ reads
\begin{equation}
n_5 = 2v \sqrt{\frac{2}{\Delta_B}} \der_t \left[{\rm Re}\left({\rm e}^{-{\rm i}\Delta_B t} \psi \right) \right] \,.
\end{equation}
By using Eq.~(\ref{comm}), we arrive at
\begin{equation}
\label{n-n_5}
[n({\bm x}), n_5({\bm y})] = -{\rm i} \frac{k}{2\pi^2} B_z \der_z \delta({\bm x} - {\bm y})\,.
\end{equation}
This is the anomalous commutator \cite{Adler:1970qb} that is 
responsible for the chiral anomaly,%
\footnote{In Ref.~\cite{Adler:1970qb}, chiral/Dirac fermions with $|k|=1$ 
are considered. The anomalous commutator~(\ref{n-n_5}) with a generic 
integer $k$ appears in a system having a monopole with charge $k$ 
in momentum space that acts as a source/sink of Berry curvature 
\cite{Son:2012wh}; see also Sec.~\ref{sec:discussion}.}
and the anomaly matching is correctly satisfied in this effective theory.

\section{Discussions}
\label{sec:discussion}
In this paper, we studied the axion electrodynamics with level $k$ in 
(3+1) dimensions in background fields as a paradigmatic example of 
topological mass generation in gapless systems. 
We have shown that this system exhibits the spontaneous breaking 
of the $\bb{Z}_k$ one-form symmetry for $k \geq  2$ even in the 
absence of the conventional topological order.

Since the low-energy effective theory of QCD coupled to QED corresponds 
to the axion electrodynamics in Eq.~(\ref{axion_elemag}) with $|k|=1$ 
as mentioned in the main text, one can ask whether the case with $|k| \geq 2$ 
can be realized in physical systems. One such possibility is the 
``axionic charge density wave'' (CDW)
\cite{Wang:2012bgb} in multi-Weyl semimetals. 
The multi-Weyl semimetal \cite{Xu, Fang} is a type of Weyl semimetal 
in which pairs of monopoles and antimonopoles with a generic integer 
charge $k$ appear in momentum space and where the chiral anomaly 
relation (\ref{anomaly}) is realized. 
In Weyl semimetals, the interaction effect may lead to the dynamical 
chiral symmetry breaking by the pairing between electrons and holes 
with opposite chiralities and, consequently, a NG mode $\phi$, which 
may be regarded as an ``axion'' field, appears \cite{Wang:2012bgb}. 
It has been recently reported that evidence of such an axionic CDW 
is experimentally observed in the Weyl semimetal (TaSe$_4$)$_2$I 
for $|k|=1$ \cite{Gooth:2019lmg}.
If the axion CDW is realized in multi-Weyl semimetals with $|k| \geq 2$, 
the effective theory for the NG mode $\phi$ coupled to dynamical 
electromagnetic fields at low energy there is the axion electrodynamics 
with level $k$.
 
In this paper we also constructed the low-energy effective theory
(\ref{EFT}) for the gapless mode with the quadratic dispersion relation,
starting from the theory (\ref{axion_elemag}) in the external magnetic
field by integrating out the gapped modes.
It would be interesting to develop a systematic construction of this
kind of low-energy effective theory based on the breaking pattern of
zero- and higher-form symmetries without referring to the details of a
microscopic theory.

\acknowledgments
R.Y.~thanks Yoshimasa Hidaka and Noriyuki Sogabe for discussions.
N.Y.~was supported by Keio Institute of Pure and Applied Sciences 
(KiPAS) project at Keio University and JSPS KAKENHI 
Grant No.~19K03852.

\end{document}